\documentclass[11pt,a4paper]{article}
\pdfoutput=1

\usepackage{jheppub}

\usepackage[utf8]{inputenc}

\usepackage{hyperref}

\usepackage{amsmath}
\usepackage{amssymb}
\usepackage{xcolor}
\usepackage{physics}
\usepackage{graphicx} 

\usepackage{pgf} 

\usepackage{tabularx}
\usepackage{multirow}
\usepackage{multicol}

\usepackage[normalem]{ulem}
\usepackage{stackengine}

\newcommand{\commie}[1]{}

\newenvironment{eqaed}
    {\begin{equation}
    \begin{aligned}}
    {\end{aligned}
    \end{equation}
    \ignorespacesafterend}

\usepackage{babel}

\makeatletter
\gdef\@fpheader{}
\makeatother

\begin{document}

\author[a]{Ivano Basile}
\author[b]{Alessia Platania} 

\affiliation[a]{Service de Physique de l'Univers, Champs et Gravitation, Universit\'{e} de Mons, Place du Parc 20, 7000 Mons, Belgium}

\affiliation[b]{Perimeter Institute for Theoretical Physics, 31 Caroline St.  N., Waterloo, ON N2L 2Y5, Canada}
  
\emailAdd{ivano.basile@umons.ac.be}
\emailAdd{aplatania@perimeterinstitute.ca}

\title{\LARGE{Asymptotic Safety: Swampland or Wonderland?}}

\begin{abstract}
     {We investigate the consequences of combining swampland conjectures with the requirement of asymptotic safety. To this end, we explore the infrared regime of asymptotically safe gravity in the quadratic one-loop approximation, and we identify the hypersurface spanned by the endpoints of asymptotically safe renormalization group trajectories. These comprise the allowed values of higher-derivative couplings as well as standard logarithmic form factors. We determine the intersection of this hypersurface with the regions of parameter space allowed by the weak-gravity conjecture, the swampland de Sitter conjecture, and the trans-Planckian censorship conjecture. The latter two depend on some order-one constants, for generic values of which we show that the overlap region is a proper subspace of the asymptotically safe hypersurface. Moreover, the latter lies inside the region allowed by the weak gravity conjecture  assuming electromagnetic duality. Our results suggest a non-trivial interplay between the consistency conditions stemming from ultraviolet completeness of the renormalization group flow, black hole physics, and cosmology.}
\end{abstract}

\maketitle


\section{Introduction}\label{sec:introduction}

The 20th century has seen the development of two pillars of modern theoretical physics: quantum field theory (QFT) and general relativity (GR). The standard model of particle physics, which successfully describes the quantum properties of the strong and electroweak interactions, is based on the former framework. However, its na\"{i}ve application to the latter yields a QFT of gravity prone to (perturbatively) non-renormalizable ultraviolet (UV) divergences~\cite{tHooft:1974toh,Goroff:1985th,Goroff:1985sz}.

Despite remarkable progress in a number of directions, the difficulties of formulating a complete theory of quantum gravity have led a considerable portion of the community to shift the focus on general, possibly model-independent lessons that could shed light on the nature of gravity at all scales. Many of these proposals, commonly dubbed ``swampland conjectures''~\cite{Vafa:2005ui} in the context of string theory, rest on considerations on black-hole physics, which often can lie entirely in the semi-classical regime where one expects low-energy effective field theory (EFT) to be a reliable description.

On the other hand, some aspects of these conjectures arose from and are tied to string theory and, in particular, its spacetime-supersymmetric incarnations. These settings are much better understood, since quantum corrections are often under quantitative control~\cite{Berglund:2005dm, Gonzalo:2018guu, Marchesano:2019ifh, Blumenhagen:2019vgj, Baume:2019sry, Palti:2020qlc, Marchesano:2021gyv, Lee:2018urn, Lee:2018spm, Lee:2019tst, Lee:2019xtm} and can sometimes even be computed exactly~\cite{Klaewer:2020lfg, Klaewer:2021vkr}. In certain settings, such as $\mathcal{N} = 2$ Calabi-Yau flux compactifications, one can even classify general families of models at once~\cite{Grimm:2018ohb, Grimm:2019wtx, Grimm:2019ixq, Gendler:2020dfp, Grimm:2020ouv, Bastian:2021eom}. In some string models with high-energy supersymmetry breaking, a variety of swampland proposals were verified~\cite{Basile:2020mpt, Basile:2021mkd}, although a construction of realistic and (meta-) stable vacua is still an open problem~\cite{Kachru:2003aw, Balasubramanian:2005zx, Koerber:2007xk, Danielsson:2009ff, Moritz:2017xto, Kallosh:2018nrk, Bena:2018fqc, Gautason:2018gln, Cordova:2018dbb, Blaback:2019zig, Hamada:2019ack, Gautason:2019jwq, Cribiori:2019clo, Andriot:2019wrs, Shukla:2019dqd, Shukla:2019akv, Cordova:2019cvf, Andriot:2020wpp, Andriot:2020vlg,Farakos:2020wfc,Gao:2020xqh,Bena:2020qpa,Bena:2020xrh,Crino:2020qwk,Dine:2020vmr,Basiouris:2020jgp,Cribiori:2020use,Hebecker:2020ejb,Andriot:2021rdy,DeLuca:2021pej,Cicoli:2021dhg,Cribiori:2021djm}. At any rate, it is paramount to understand the consequences of general consistency conditions outside of the specific contexts arising from (supersymmetric) string compactifications, although parametric control is most likely going to be problematic~\cite{Dine:1985he,Bena:2018fqc,Gao:2020xqh,Bena:2020xrh,Dine:2020vmr} due to unknown or uncalculable corrections. To wit, efforts to test swampland proposals have almost entirely focused on supersymmetric settings, and in particular on stringy constructions. In order to shed light on whether they only encode a ``string lamppost principle''~\cite{Montero:2020icj} or they hold in more generality, it is important to extend their exploration to a broader class of quantum gravity models. On the other hand, the more stringent and well-grounded swampland proposals, such as the ``no global symmetries''~\cite{Misner:1957mt, Polchinski:2003bq, Banks:2010zn, Harlow:2018tng} and weak gravity~\cite{Arkani-Hamed:2006emk} conjectures, could help guide the search for asymptotic safety, which is currently faced with the daunting prospect of navigating ever-larger theory spaces~\cite{Benedetti:2010nr, Knorr:2021slg}.

A concrete problem that can be phenomenologically relevant in the near future is constraining the possible values of the Wilson coefficients of a curvature expansion of the gravitational EFT. In particular, one can expect that generic detectable leading-order effects of quantum gravity be encoded in the coefficients of the quadratic curvature invariants, which we shall discuss in detail in the following, or in some specific non-local form factors~\cite{Belgacem:2017cqo,Knorr:2018kog}. Some efforts in this direction have been made using the S-matrix bootstrap~\cite{Guerrieri:2021ivu, Caron-Huot:2021rmr}, finding compelling agreement with the parameter space allowed by string theory. In the EFT framework, the problem has also been investigated via positivity bounds~\cite{deRham:2017zjm, deRham:2017xox, deRham:2018qqo, DeRham:2018bgz, Alberte:2019lnd, Alberte:2019xfh, Alberte:2019zhd, Alberte:2020bdz, Alberte:2020jsk, deRham:2021fpu}.

In this paper we approach this issue from a novel direction, studying the constraints coming from swampland conjectures together with the consistency conditions required by the existence of a UV fixed point of the gravitational renormalization group (RG) flow\footnote{See also~\cite{deAlwis:2019aud} for related discussions on the weak gravity conjecture in the context of asymptotically safe gravity.}. The latter scenario has been termed ``asymptotic safety'', in analogy with asymptotic freedom as a particular case. This idea, originally due to Weinberg~\cite{1976W}, would imply that the Wilson coefficients of the IR effective action stem from a UV-complete RG trajectory, which in turn would be determined by a finite number of relevant deformations from the fixed point. Recently, this area of research has witnessed considerable development of the theoretical framework to investigate RG flows beyond perturbation theory~\cite{Dupuis:2020fhh} and finding evidence for the existence of the Reuter fixed point in a variety of different approximation schemes~\cite{Souma:1999at,Lauscher:2002sq,Litim:2003vp,Codello:2006in,Machado:2007ea,Benedetti:2009rx,Dietz:2012ic,Dona:2013qba,Eichhorn:2013xr,Dona:2014pla,Christiansen:2014raa,Falls:2014tra,Christiansen:2015rva,Meibohm:2015twa,Oda:2015sma,Dona:2015tnf,Biemans:2016rvp,Eichhorn:2016esv, Dietz:2016gzg, Falls:2016msz,Gies:2016con,Biemans:2017zca,Christiansen:2017cxa,Hamada:2017rvn,Platania:2017djo,Falls:2017lst,deBrito:2018jxt,Eichhorn:2018nda,Eichhorn:2019yzm,deBrito:2019umw,Knorr:2021slg} (see also~\cite{Donoghue:2019clr,Bonanno:2020bil} for critical assessments on the status of the field and its open questions). Possible implications of asymptotically safe gravity in astrophysics and cosmology (see~\cite{Bonanno:2017pkg,Platania:2020lqb} for reviews) have been investigated using simplified models~\cite{Bonanno:2006eu,Falls:2012nd,torres15,Koch:2015nva,Bonanno:2015fga,Bonanno:2016rpx,Kofinas:2016lcz,Falls:2016wsa,Bonanno:2016dyv,Bonanno:2017gji,Bonanno:2017kta,Bonanno:2017zen,Bonanno:2018gck,Liu:2018hno,Majhi:2018uao,Anagnostopoulos:2018jdq,Adeifeoba:2018ydh,Pawlowski:2018swz,Gubitosi:2018gsl,Platania:2019qvo,Platania:2019kyx,Bonanno:2019ilz,Held:2019xde} and more elaborate computations~\cite{Bosma:2019aiu}, leading to the tentative conclusions that black-hole and cosmological singularities could be resolved by quantum effects, and that the nearly scale-invariant  cosmological power spectrum could arise naturally from a nearly scale-invariant  asymptotically safe regime.

In this paper we shall propose a concrete method to extract the allowed region of IR parameters from the RG flow of asymptotically safe trajectories. 
In particular, we shall focus on the simpler case of the one-loop approximation in quadratic gravity~\cite{Codello:2006in,Niedermaier:2009zz, Niedermaier:2010zz} in order to test our construction and provide a proof of principle of our idea. We will show that the IR limit of asymptotically safe trajectories falls inside the region allowed by the weak gravity conjecture and electromagnetic duality, and display a non-trivial intersection with the one allowed by the de Sitter and trans-Planckian censorship bounds.

The contents of this paper are organized as follows. In sect.~\ref{sec:swampland} we provide a brief overview of swampland conjectures, focusing on the weak gravity conjecture, the de Sitter conjecture and the trans-Planckian censorship conjecture, since they entail the most relevant bounds for our subsequent analysis. In sect.~\ref{sec:one-loop} we describe in detail the one-loop approxima\-tion to quadratic gravity that we employ as testing grounds, and our method of extracting the physical IR Wilson coefficients. The resulting effective action turns out to contain non-local form factors. In sect.~\ref{sec:results} we collect our results: in sect.~\ref{sec:IR_space} we present the allowed region of parameter space that we found, which spans a plane in the three-dimensional space of dimensionless IR parameters, and in sect.~\ref{sec:wgc_constraints} and sect.~\ref{sec:dsc_constraints} we study the constraints stemming from the swampland conjectures discussed in sect.~\ref{sec:wgc_intro} and sect.~\ref{sec:dsc_tcc_intro} respectively. In sect.~\ref{sec:intersections} we discuss and display the intersection of all regions. We conclude with a summary and some perspectives in sect.~\ref{sec:conclusions}.

\section{An overview of swampland conjectures}\label{sec:swampland}

As we have anticipated in the introduction, swampland conjectures are proposals that ought to rule out EFTs of gravity that do not admit UV completions~\cite{Vafa:2005ui}. These conjectures are generally motivated in part by purely low-energy considerations, stemming from black-hole physics or inflation, but they also arise from detailed investigations of string-theoretic settings, where generally one has more control over corrections and patterns can be corrobo\-rated across families of EFTs. The latter approach has led some to describe a ``lamppost'' effect~\cite{Montero:2020icj}, whereby only settings that are somewhat under control can be investigated and thus it is unclear to which extent the resulting conclusions can be generalized. Furthermore, while at least minimal supersymmetry is generally unbroken in order to retain computational control, recent considerations~\cite{Cribiori:2021gbf,Castellano:2021yye} point to a tension between low-energy supersymme\-try breaking\footnote{Nevertheless, scenarios with high-energy supersymmetry breaking have been investigated in the context of the swampland~\cite{Bonnefoy:2018tcp,Basile:2020mpt,Basile:2021mkd}. See~\cite{Mourad:2017rrl, Basile:2021vxh, Mourad:2021lma} for recent reviews.} and the consistency of the EFT. As we have discussed in the preceding section, one of the motivations behind this work is indeed to go beyond the usual settings, seeking lessons for other approaches to quantum gravity.

Since its inception, the swampland program aims to describe the boundary between the landscape of consistent gravitational EFTs with a growing number of proposed criteria\footnote{See~\cite{Palti:2019pca, vanBeest:2021lhn} for reviews.}, numerous relations among which~\cite{Andriot:2020lea, Lanza:2020qmt} point to a deeper underlying principle. In particular, connections between the distance conjecture~\cite{Ooguri:2006in, Ooguri:2018wrx} and string dualities suggest that an organizing principle for these consistency criteria in the IR be related to a non-perturbative UV formulation of quantum gravity. Furthermore, as we shall see in the following, swampland considerations have provided intriguing clues toward a number of phenomenological puzzles~\cite{Grana:2021zvf}. 

In this paper we shall focus on some conjectures which can provide bounds for the Wilson coefficients of the gravitational EFT. In particular,

\begin{itemize}
    \item The \emph{weak gravity conjecture} (WGC)~\cite{Arkani-Hamed:2006emk} relates the mass and charge of light states and black holes;
    \item The \emph{de Sitter conjecture} (dSC)~\cite{Obied:2018sgi}, along with its refined versions~\cite{Ooguri:2018wrx,Garg:2018reu,Andriot:2018mav}, constrains the behavior of scalar potentials and their derivatives, leading to an obstruction to the existence of de Sitter vacua that is $\mathcal{O}(1)$ in Planck units;
    \item The \emph{trans-Planckian censorship conjecture} (TCC)~\cite{Bedroya:2019snp, Brandenberger:2021pzy} constrains sub-Planckian cosmological perturbations to remain sub-Planckian across inflation, and leads to bounds on the lifetime of metastable de Sitter configurations as well as on the $\mathcal{O}(1)$ parameter that appears in the dSC, at least in asymptotic regions of field space.
\end{itemize}

In light of the latter consideration, for the purposes of this paper in the following we shall investigate the consequences of the TCC on Starobinsky-like inflationary potentials as a special case of the dSC. Indeed, we shall restrict ourselves to the asymptotic region of field space corresponding to small curvatures in Planck units, where the TCC could provide a dSC bound with a specific $\mathcal{O}(1)$, as we shall see below.

\subsection{Weak gravity conjecture and black holes}\label{sec:wgc_intro}

Let us begin reviewing some features of the (electric) WGC, referring the reader to~\cite{Palti:2019pca,  vanBeest:2021lhn} for more details. In its most basic form, it states that in a consistent EFT of gravity coupled to a $U(1)$ gauge field there exists a state whose mass $m$ is lower than its charge $q$ in Planck units. In four dimensions, the bound for charged particles reads
\begin{eqaed}\label{eq:wgc_basic}
    \frac{m}{M_\text{Pl}} \leq \mathcal{O}(1) \, q \, ,
\end{eqaed}
where the model-dependent $\mathcal{O}(1)$ constant is $\frac{1}{\sqrt{2}}$ in Einstein-Maxwell theory.

Among various motivations and evidence gathered in the literature, the WGC is grounded in black-hole physics from the requirement that charged, extremal black holes be able to decay, lest protected by a symmetry (such as supersymmetry, in the case of BPS-saturated states). The rationale behind this lies in avoiding remnants while keeping the black hole from violating the extremality bound, since a violation of either would presumably lead to consistency issues potentially within the EFT regime~\cite{Giddings:1992hh, Susskind:1995da, Arkani-Hamed:2006emk}. For charged black holes of mass $M$ and charge $Q$, this requirement translates into
\begin{eqaed}\label{eq:wgc_bhs}
    \frac{M}{Q} \geq \left(\frac{M}{Q}\right)_\text{extremal} \, ,
\end{eqaed}
where the latter is generally an $\mathcal{O}(1)$ constant\footnote{For Einstein-Maxwell theory, the extremality bound reads $\frac{M}{Q} \geq \sqrt{2} \, M_\text{Pl}$.}. However, higher-curvature corrections could potentially spoil this condition even for macroscopic black holes, provided they are sufficiently close to extremality. Writing the leading quartic corrections according to~\cite{Kats:2006xp}
\begin{eqaed}\label{eq:eft_corr}
    \Delta \mathcal{L} = c_1 \, R^2 + c_2 \, R_{\mu \nu}R^{\mu \nu} + c_3 \, R_{\mu \nu \rho \sigma} R^{\mu \nu \rho \sigma} \, ,
\end{eqaed}
the resulting bounds for the corresponding Wilson coefficients $c_i$ comprise a family of inequalities for linear combinations of the $c_i$, parametrized by the extremality parameter of the black hole~\cite{Arkani-Hamed:2006emk,Kats:2006xp,Cheung:2018cwt,Charles:2017dbr,Hamada:2018dde,Charles:2019qqt}. The extremality bound in general now takes the form
\begin{eqaed}\label{eq:wgc_bhs_hd}
    \frac{M}{Q} \geq \left(\frac{M}{Q}\right)_\text{extremal} \left(1 - \frac{\Delta}{M^2} \right) \, ,
\end{eqaed}
where the linear combination $\Delta$ of Wilson coefficients is to be non-negative in order for the WGC to hold, and is proportional to the coefficient $c_2 + 4c_3$ of the Weyl-squared term~\cite{Kats:2006xp,Charles:2019qqt}.

The leading order contributions to $\Delta$ comprise not only the Wilson coefficients in the effective action of eq.~\eqref{eq:eft_corr}, but also the Wilson coefficients that involve the $U(1)$ gauge field. It has been recently shown~\cite{Cano:2021tfs} that, \emph{assuming invariance under electromagnetic duality}, higher-curvature corrections up to sextic order can be written in terms of purely gravitational terms, up to field redefinitions. Let us stress that our aim is to intersect swampland bounds with the constraints provided by asymptotic safety, and the technical obstacles to compute its consequences for quartic electromagnetic couplings in gravity, which would entail involved FRG computations along the lines of~\cite{Knorr:2021slg}, compel us to focus on the duality-invariant scenario of~\cite{Cano:2021tfs}, which at any rate appears intriguing on its own\footnote{Another instance of the interplay between duality and the WGC has been studied in~\cite{Loges:2020trf}.}. Moreover, the electromagnetic couplings do not run under the RG flow at one loop because of tree-level duality~\cite{Charles:2017dbr,Charles:2019qqt}. This has been used to argue that the low-energy behavior of the correction $\Delta$ to the extremality bound is dominated by the Weyl anomaly coefficients that drive the running of the $c_i$, and in particular of the Weyl-squared Wilson coefficient $c_2 + 4c_3$, as we shall see in the following. However, in the present case we would like to constrain the physical parameters built out of the Wilson coefficients and the Planck scale, in order to compare the resulting bounds with the constraints of asymptotic safety. We shall describe the procedure in detail in the following section.

\subsection{de Sitter and trans-Planckian censorship}\label{sec:dsc_tcc_intro}

Let us now move on to discuss the dSC and the TCC. The former quantifies an obstruction to the existence of de Sitter vacua, in the form of a bound for the (field-space gradient of the) scalar potential $V(\phi)$. Indeed, since in this setting de Sitter vacua would arise as positive-energy critical points of $V$, a natural bound that would prevent these takes the form
\begin{eqaed}\label{eq:dsc_basic}
    M_\text{Pl}\, \abs{\nabla V} \geq c V 
\end{eqaed}
for field ranges
\begin{eqaed}\label{eq:field_range}
    \Delta \phi \lesssim f \, M_\text{Pl} \, ,
\end{eqaed}
where $c \, , \, f > 0$ are (\emph{a priori} model-dependent) constants. Within the EFT framework, their natural values are $\mathcal{O}(1)$, indicating that the obstruction is tied to the expected cutoff of the EFT. However, one is readily confronted with a tension between the bound in eq.~\eqref{eq:dsc_basic} and slow-roll inflation~\cite{Garg:2018reu, Kinney:2018nny}\footnote{See also~\cite{Rudelius:2019cfh,Rudelius:2021oaz,Chojnacki:2021fag,Jonas:2021xkx} for discussions on eternal inflation and the swampland.}, leading to refinements involving the Hessian matrix of the potential~\cite{Garg:2018reu,Ooguri:2018wrx}. In particular, whenever the bound of eq.~\eqref{eq:dsc_basic} would be violated, the matrix
\begin{eqaed}\label{eq:ref_dsc}
    M_\text{Pl}^2\, \text{Hess}(V) + \, c' \, V
\end{eqaed}
would be negative semidefinite, with $c' > 0$ another $\mathcal{O}(1)$ constant. Further refinements were proposed in~\cite{Andriot:2018mav}, but in our setting we shall find that the first bound of eq.~\eqref{eq:dsc_basic} is sufficient, since it encompasses eq.~\eqref{eq:ref_dsc} in the regions of parameter space that we are concerned with.

On the other hand, the TCC surmises that sub-Planckian quantum fluctuations in the early universe at initial time $t_i$ never grow macroscopic at a final time $t_f$. In particular, they ought to never cross the Hubble horizon and freeze. This requirement can be formulated, in terms of the scale factor $a(t)$ and the corresponding Hubble parameter $H$, by~\cite{Bedroya:2019snp, Brandenberger:2021pzy}
\begin{eqaed}\label{eq:tcc_basic}
    \frac{a(t_f)}{a(t_i)} \lesssim \frac{M_\text{Pl}}{H(t_f)} \, ,
\end{eqaed}
again up to an $\mathcal{O}(1)$ constant. An intriguing consequence of eq.~\eqref{eq:tcc_basic} is that de Sitter configurations are not prohibited, but they are metastable with a lifetime $T$ bounded by
\begin{eqaed}\label{eq:tcc_life}
    T \lesssim \frac{1}{H} \, \log \frac{M_\text{Pl}}{H} \, ,
\end{eqaed}
of the order of a trillion years. This results points to a possible resolution of the coincidence problem in this setting.

The most relevant consequence of the TCC for the purposes of this paper is that, in the presence of a scalar potential, it leads to a bound of the form of eq.~\eqref{eq:dsc_basic} with
\begin{eqaed}\label{eq:tcc_c1}
    c = \frac{2}{\sqrt{(d-1)(d-2)}}
\end{eqaed}
in $d$ spacetime dimensions, at least in asymptotic regimes of field space. In the present setting, the scalar potential arises from the quadratic curvature terms, and the corresponding asymptotic regime for gravitational field fluctuations is that of small curvatures~\cite{Lust:2019zwm}. This regime is mapped to a neighbourhood of the origin in the inflaton description. For generic curvatures, one expects that both the purely gravitational description and the inflaton description be modified, including the geometry of field space. Nevertheless, since our current setup does not allow for precise quantitative bounds, we shall henceforth take eq.~\eqref{eq:tcc_c1} simply as a reference point around which to study the more general bound of eq.~\eqref{eq:dsc_basic}. Let us also remark that this value appears in a number of related swampland bounds~\cite{Andriot:2020lea} and is well-behaved under dimensional reduction~\cite{Rudelius:2021oaz}, and thus it may play a more prominent role in the story. At any rate, it would be interesting to explore the more direct implications of eq.~\eqref{eq:tcc_basic} studying cosmological solutions or exploring the considerations of~\cite{Bedroya:2019tba, Brandenberger:2021pzy} within our setup.

\section{One-loop RG flow in quadratic gravity}\label{sec:one-loop}

Let us now discuss the concrete setting in which we shall compute the possible values of the Wilson coefficients of the effective gravitational action. In this work we focus on the quadratic truncation\footnote{Let us remark that here ``quadratic'' refers to the order in the curvatures. In terms of derivatives, the action in eq.~\eqref{eq:quadratic_lagrangian} is quartic.}, in the one-loop approximation. In Euclidean signature, the Lagrangian pertaining to the full quadratic truncation reads
\begin{eqaed}\label{eq:quadratic_lagrangian}
    \mathcal{L}=\frac{2\Lambda-R}{16\pi G}+\frac{1}{2\lambda}\,C^2-\frac{\omega}{3\lambda}R^2+\frac{\theta}{\lambda}E \, ,
\end{eqaed}
where $C^2 \equiv C_{\mu\nu\rho\sigma}C^{\mu\nu\rho\sigma}$ is the square of the Weyl tensor, $E$ is the Gauss-Bonnet density and and the Wilson coefficients
\begin{eqaed}\label{eq:wilson_coeffs}
    g_C \equiv \frac{1}{2\lambda} \, , \qquad
    g_R \equiv - \, \frac{\omega}{3\lambda}
\end{eqaed}
can be related to the $c_i$ coefficients in eq.~\eqref{eq:eft_corr}. Indeed, since
\begin{eqaed}\label{eq:weyl_gb}
    C^2&=R_{\mu\nu\rho\sigma}R^{\mu\nu\rho\sigma}-2R_{\mu\nu}R^{\mu\nu}+\frac{R^2}{3}\,,\\
    E&=R_{\mu\nu\rho\sigma}R^{\mu\nu\rho\sigma}-4R_{\mu\nu}R^{\mu\nu}+R^2\,,
\end{eqaed}
the $c_i$ are related to the couplings in eq.~\eqref{eq:quadratic_lagrangian} according to
\begin{eqaed}\label{eq:couplings_relation}
    c_1&=\frac{1}{6\lambda}-\frac{\omega}{3\lambda}+\frac{\theta}{\lambda} \, ,\\
    c_2&=-\frac{1}{\lambda}-\frac{4\theta}{\lambda} \,,\\
    c_3&=\frac{1}{2\lambda}+\frac{\theta}{\lambda} \,.
\end{eqaed}
While this setup holds in general spacetime dimensions $d$, we now restrict to $d = 4$. The one-loop beta functions of the couplings of eq.~\eqref{eq:quadratic_lagrangian} read~\cite{Codello:2006in,Niedermaier:2009zz, Niedermaier:2010zz}
\begin{eqaed}\label{eq:betas}
\beta_{\widetilde{\Lambda}} & =-\,2 \widetilde{\Lambda}+\frac{1}{(4 \pi)^{2}}\left[\frac{1+20 \omega^{2}}{256 \pi \widetilde{G} \omega^{2}} \lambda^{2}+\frac{1+86 \omega+40 \omega^{2}}{12 \omega} \lambda \widetilde{\Lambda}\right] \\
& \quad -\frac{1+10 \omega^{2}}{64 \pi^{2} \omega} \lambda +\frac{2 \widetilde{G}}{\pi}-\frac{83+70 \omega+8 \omega^{2}}{18 \pi} \widetilde{G} \widetilde{\Lambda} \, , \\
\beta_{\widetilde{G}} & =2 \widetilde{G}-\frac{1}{(4 \pi)^{2}} \frac{3+26 \omega-40 \omega^{2}}{12 \omega} \lambda \widetilde{G}-\frac{83+70 \omega+8 \omega^{2}}{18 \pi} \widetilde{G}^{2} \, ,
\\
\beta_{\lambda} & =-\frac{1}{(4 \pi)^{2}} \frac{133}{10} \lambda^{2} \, , \\
\beta_{\omega} & =-\frac{1}{(4 \pi)^{2}} \frac{25+1098 \,\omega+200 \,\omega^{2}}{60} \lambda \, , \\
\beta_{\theta} & =\frac{1}{(4 \pi)^{2}} \frac{7\,(56-171\, \theta)}{90} \lambda
\end{eqaed}
where $\widetilde{G}_k=G_k\,k^2$ and $\widetilde{\Lambda}_k=\Lambda_k\,k^{-2}$
are the dimensionless Newton coupling and cosmological constant respectively, and we have suppressed the subscript $k$ in eq.~\eqref{eq:betas} for the sake of clarity.

\begin{figure}
\centering\includegraphics[scale=0.7]{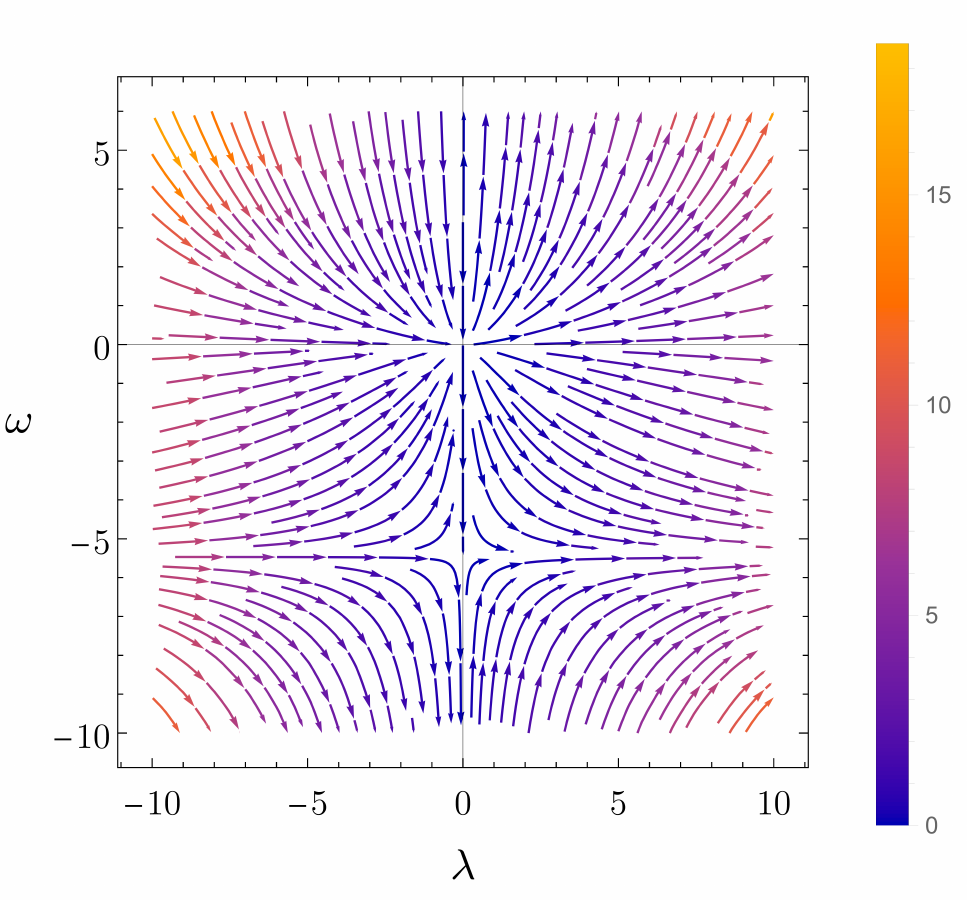}
\caption{Flow of one-loop quadratic gravity in the $(\omega,\lambda)$ subspace. The arrows point toward the IR. In this setting, two non-trivial fixed points are present, and the Reuter fixed point is the one with the smaller absolute value~\cite{Codello:2008vh}. \label{fig:flowlambdaomega}}
\end{figure}

In our setting, the flow of the (classically) marginal couplings $\lambda$, $\omega$ and $\theta$ is decoupled from that of the Einstein-Hilbert couplings. Out of the UV fixed points
\begin{eqaed}\label{eq:fixed_points}
    \lambda_\ast = 0 \, , \qquad \omega_\ast = \omega_\pm \equiv \frac{-549 \pm 7\sqrt{6049}}{200} \, , \qquad \theta_\ast = \frac{56}{171} \, ,
\end{eqaed}
UV completeness selects $\omega_\ast = \omega_+ \approx - \, 0.023$~\cite{Codello:2008vh}, which the solutions approach as the RG time\footnote{Note that, since we are interested in the IR regime, our convention for the RG time is such that $t \to +\infty$ in the IR.} $t \equiv \log\frac{k_0}{k} \to -\infty$, as is apparent from fig.~\ref{fig:flowlambdaomega}. Let us remark that this fixed point is asymptotically safe, \emph{i.e.} at least one coupling is not asymptotically free~\cite{Codello:2008vh, Niedermaier:2009zz, Niedermaier:2010zz, Groh:2011vn}. Indeed, the critical exponents of $G$ and $\Lambda$ are 2 and 4\footnote{Although 2 and 4 are not the canonical mass dimensions of $G$ and $\Lambda$, they are the canonical dimension of the couplings $1/G$ and $\Lambda/G$ that multiply the operators $\sqrt{-g}$ and $\sqrt{-g} R$. This occurs because the transformation between these couplings is non-singular, as explained in~\cite{Percacci:2007sz}. On the other hand, at the Gaussian fixed point the transformation between the couplings is singular, and the dimensions change accordingly.}, while in the IR they become the canonical -2 and 2 respectively~\cite{Litim:2012vz}. The fact that all couplings are attracted to the fixed point in the UV~\cite{Codello:2008vh} is instead an artifact of the one-loop approximation. Indeed, more sophisticated FRG computations yield a fixed point with a three-dimensional critical surface~\cite{Benedetti:2009rx}.

\begin{figure}[t!]
\centering\includegraphics[scale=0.4]{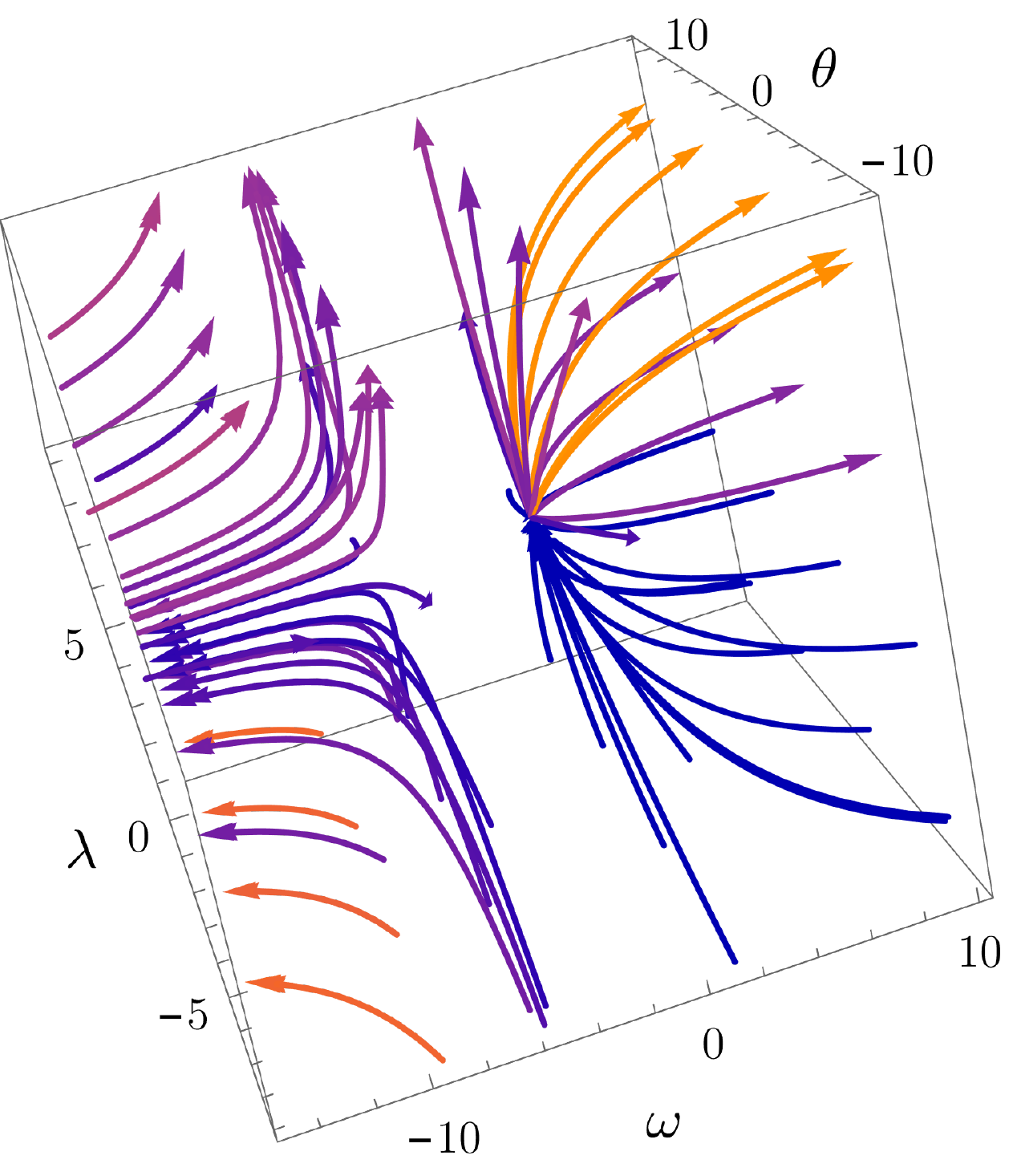}\\\belowbaseline[0pt]{
\includegraphics[scale=0.45]{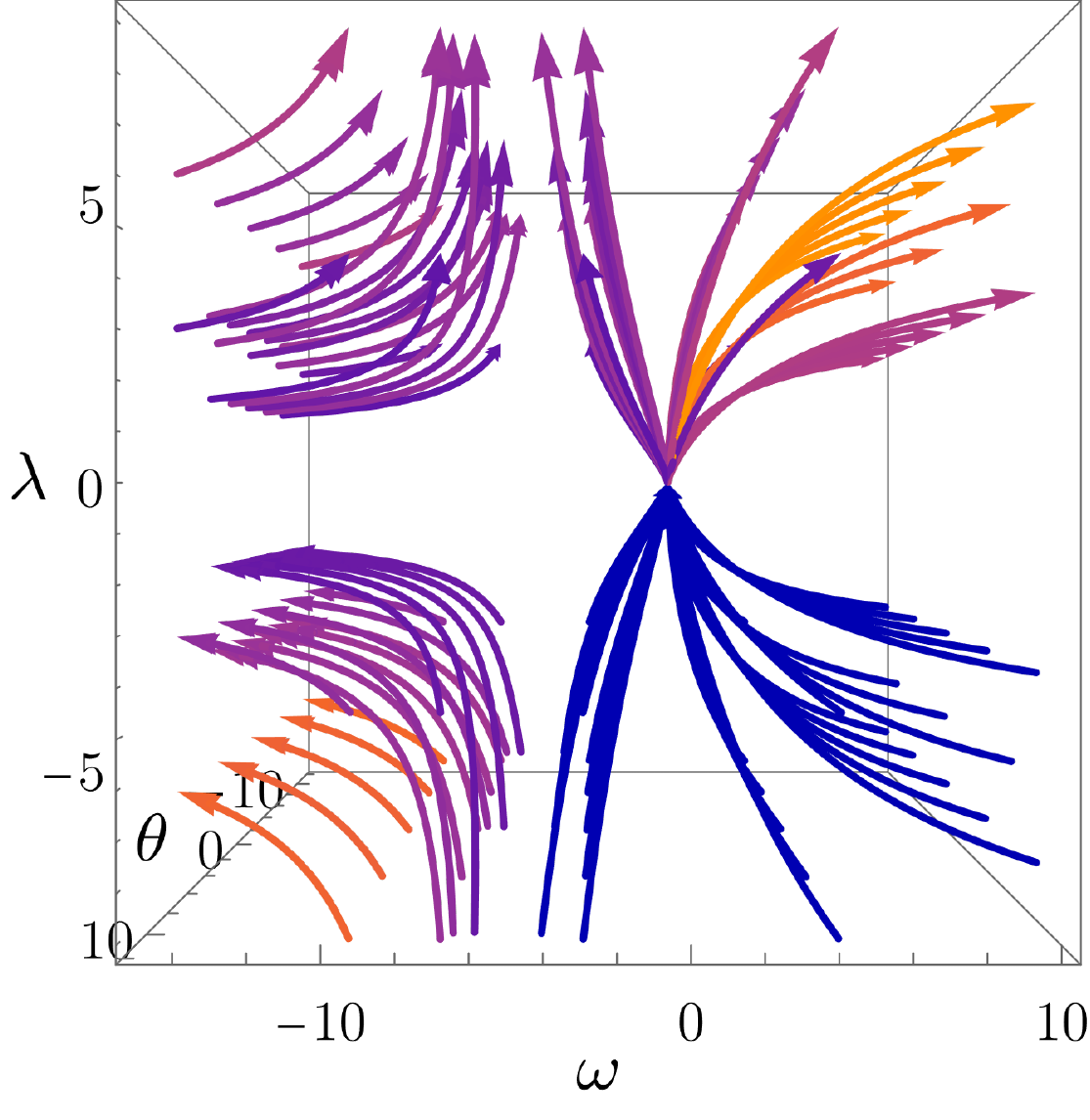}}~$\qquad$\belowbaseline[0pt]{\includegraphics[scale=0.47]{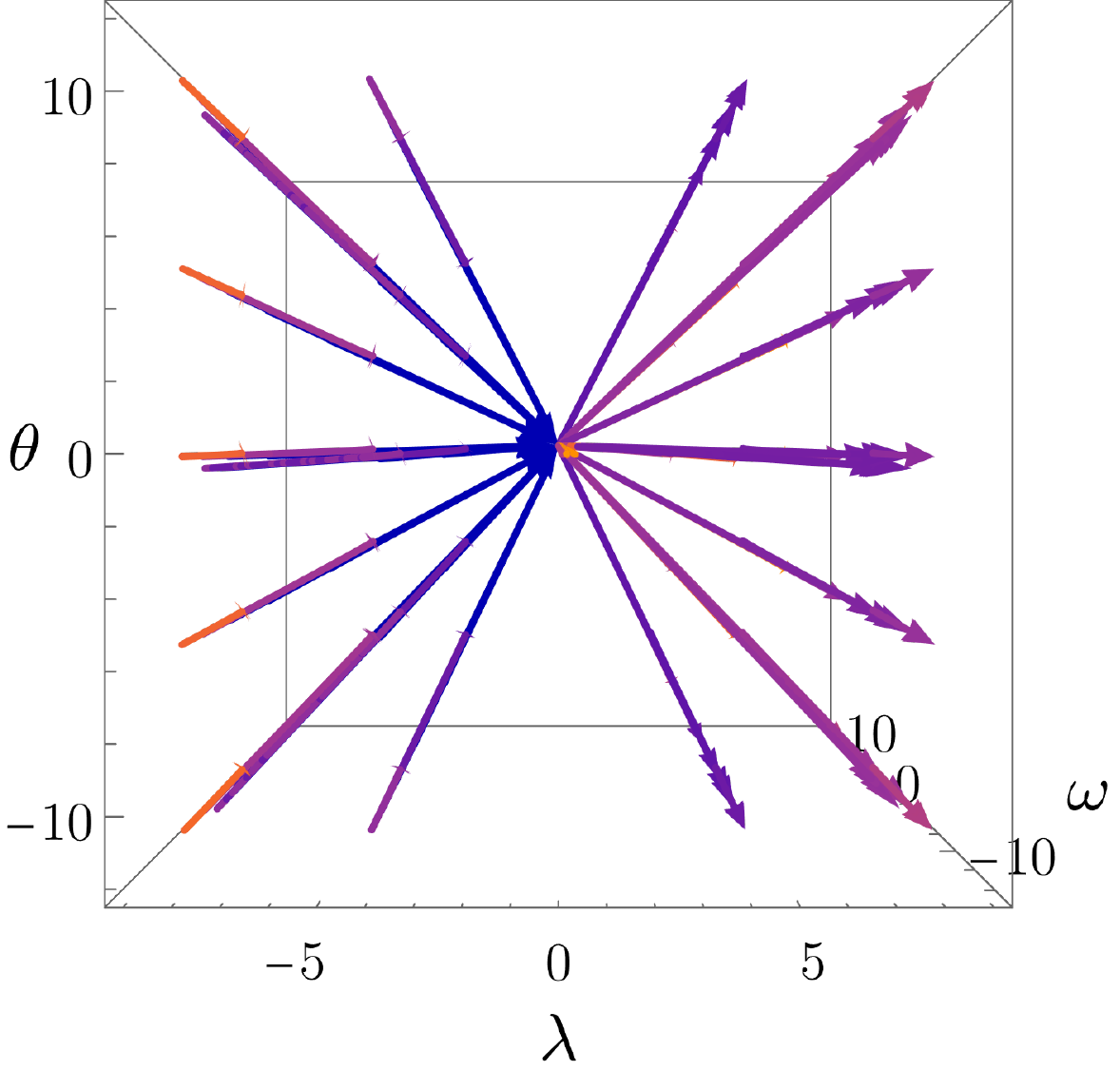}}
\caption{Flow of the classically marginal couplings $(\omega,\lambda,\theta)$ in one-loop quadratic gravity. The arrows point towards the IR, and different viewpoints are shown to better visualize the flow. The color coding of the arrows is identical to that of fig.~\ref{fig:flowlambdaomega}.\label{fig:flowMC}}
\end{figure}

\begin{figure}
\centering\includegraphics[scale=0.68]{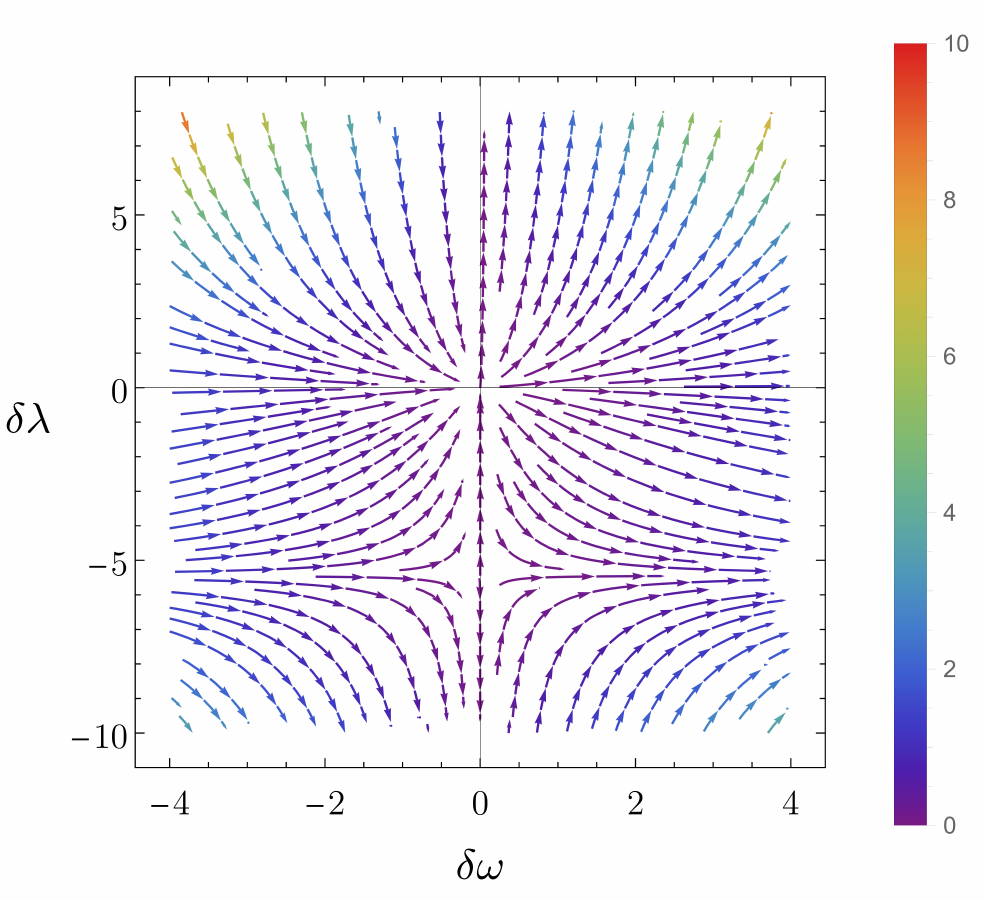}\includegraphics[scale=0.68]{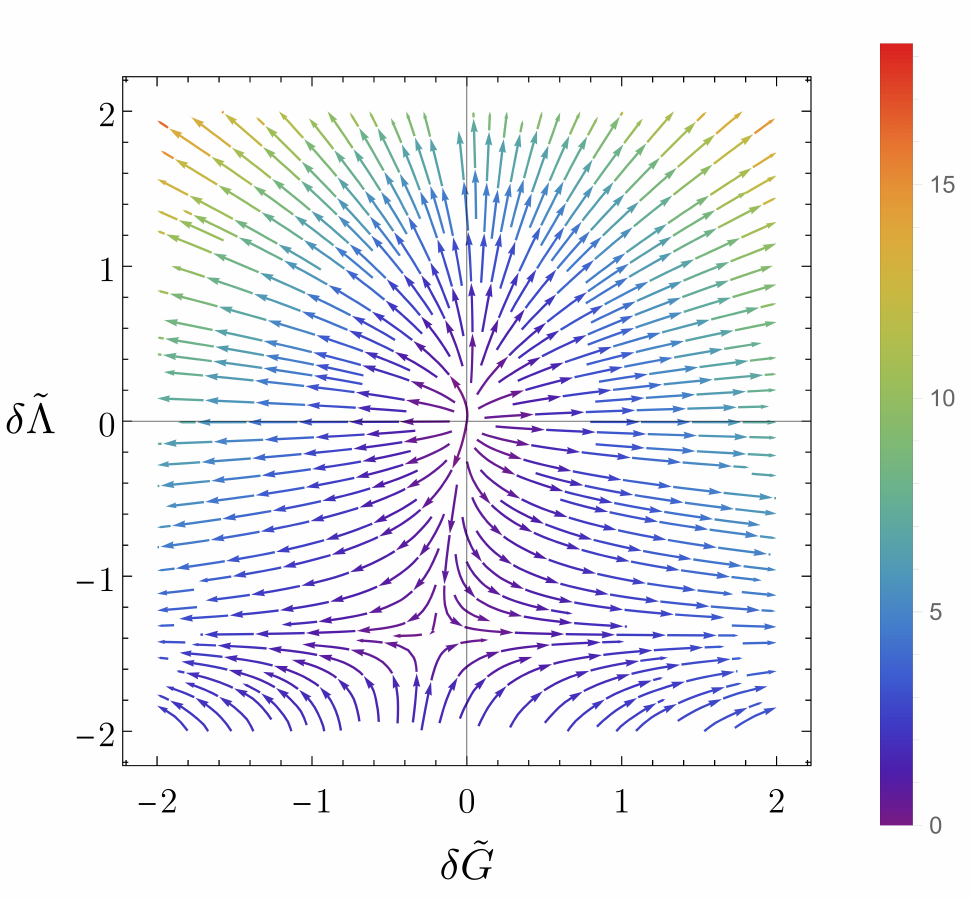}
\caption{Flow of the relevant deformations from the fixed point. The left panel depicts the flow in the $(\omega,\lambda)$ subspace. The right panel depicts the flow in the $(G, \Lambda)$ subspace where the classically marginal couplings have been set to the UV fixed point. The arrows indicate the the flow from the UV to the IR.\label{fig:flow-reldeformations}}
\end{figure}

The flow can be solved analytically in terms of the deformations $\delta \lambda \, , \, \delta \omega$ from the UV fixed point, and yields the closed-form solution
\begin{eqaed}\label{eq:marginal_flow}
    \lambda(t) & = \frac{\delta \lambda}{1 - \frac{133}{160\pi^2} \, \delta \lambda \, t} \,,
    \\
    \omega(t) & = \frac{\omega_- - \, \omega_+ \, \left(1 + \frac{\Delta}{\delta \omega}\right) \left(1 - \, \frac{133}{160\pi^2} \, \delta \lambda \, t \right)^{\frac{7\sqrt{6049}}{399}}}{1 - \, \left(1 + \frac{\Delta}{\delta \omega}\right) \left(1 - \, \frac{133}{160\pi^2} \, \delta \lambda \, t \right)^{\frac{7\sqrt{6049}}{399}}} \,,
    \\
    \theta(t) & = \frac{56}{171} + \frac{\delta \theta}{1 - \frac{133}{160\pi^2} \, \delta \lambda \, t} \,,
\end{eqaed}
where $\Delta \equiv \omega_+ - \omega_-$. The vector field generating this flow is displayed in fig.~\ref{fig:flowlambdaomega} in the $(\omega,\lambda)$ subspace and in fig.~\ref{fig:flowMC} with various 3D plots. Let us observe that the UV completeness of the trajectory requires $\delta \lambda > 0$, and that the IR flow ends at $t = t_\text{IR} \equiv \frac{160\pi^2}{133 \delta \lambda}$. However, since $\delta \lambda \ll 1$ this RG time is parametrically large, and one can reliably extract the perturbative IR behavior for the Wilson coefficients. Furthermore, reaching a physical IR regime with $\widetilde{G} \to 0^+$ requires that $\delta \widetilde{G} \, , \, \delta \omega < 0$, in order that their flows remain between the UV and IR fixed-point values avoiding runaway. The flow of the relevant deformations from the fixed point is shown in Fig.~\ref{fig:flow-reldeformations}.

Substituting the expressions of eq.~\eqref{eq:marginal_flow} in eq.~\eqref{eq:betas}, one can then solve the remaining flow equations numerically varying the initial conditions, or, equivalently, the deformations $\delta \omega, \delta \lambda, \delta \widetilde{G} , \delta \widetilde{\Lambda}$ from the UV fixed point. The RG flow then drives the running couplings to the weakly coupled IR, where the running couplings $g_C$ and $g_R$, defined in eq.~\eqref{eq:wilson_coeffs}, behave logarithmically (linearly in $t$) as $t \to t_\text{IR}^-$. This result is consistent with perturbative computations, and the resulting asymptotic expressions read
\begin{eqaed}\label{eq:log_running}
    g_C(t) & \sim \frac{1}{2\delta \lambda} - \frac{133}{320\pi^2} \, t \, , \\
    g_R(t) & \sim - \, \frac{\omega_-}{3\delta \lambda} + \frac{133}{480\pi^2} \, \omega_- \, t \, .
\end{eqaed}
In order to extract the physical IR parameters, we shall identify the (square of the) RG scale $k^2$ with the covariant Laplacian/d'Alembertian $\Box$. In order to eliminate the arbitrary reference scale $k_0$ that defines the initial condition for the RG flow, one can express every quantity in units of the IR Planck mass\footnote{Notice that our convention for the Planck mass differs from the more widespread ``reduced'' Planck mass $\widehat{M}_\text{Pl}^{-2} = 8\pi G$.} $M_\text{Pl}^{-2} = G$. To this end, since $e^{2t} \, \widetilde{G}(t) \to \widetilde{G}_0$ tends to a constant in the IR, one can evaluate the running Wilson coefficients of eq.~\eqref{eq:log_running} replacing $t \to - \, \frac{1}{2} \log \widetilde{G}(t)$, so that
\begin{eqaed}\label{eq:t_planck_sub}
    \log \frac{M_\text{Pl}}{k} & = \log \frac{M_\text{Pl}}{k_0} + t \\
    & \sim - \, \frac{1}{2} \log \widetilde{G}(t) \\
    &  \sim - \, \frac{1}{2} \log \widetilde{G}_0 + t \, .
\end{eqaed}
Since $\widetilde{G}_0$ can be extracted from the numerical solution of eqs.~\eqref{eq:betas}, identifying 
\begin{eqaed}\label{eq:form_factor_sub}
    \log \frac{k^2}{M_\text{Pl}^2} \longrightarrow \log \frac{\Box}{M_\text{Pl}^2} \, ,
\end{eqaed}
according to the preceding considerations, one can reconstruct an effective action of the form
\begin{eqaed}\label{eq:form_factor_eft}
    \Gamma = \int d^4x \sqrt{g} \, \left(\frac{2\Lambda - R}{16\pi G} + g_C \, C^2 + g_R \, R^2 + b_C \, C \log \frac{\Box}{M_\text{Pl}^2} \, C + b_R \, R \log \frac{\Box}{M_\text{Pl}^2} \, R \right) \, , 
\end{eqaed}
where Weyl-tensor contraction is understood. The appearance of non-local form factors resonates with the considerations in~\cite{Knorr:2019atm, Draper:2020bop, Draper:2020knh}. While we shall neglect them in the following, the presence of form factors of this type seems largely consistent with preceding results~\cite{Riegert:1984kt, Deser:1996na, Erdmenger:1996yc, Erdmenger:1997gy, Deser:1999zv, Bautista:2017enk} (see also~\cite{Donoghue:2015nba} for a discussion of logarithmic form factors). Note that, despite their behavior at low energies, one expects a resummation of such non-local form factors to yield a result that is both well-defined and subleading in the IR compared to the local terms~\cite{Draper:2020bop}\footnote{One exception could be a non-local form factor of the type $\sim 1/\Box$, as discussed in~\cite{Belgacem:2017cqo,Knorr:2018kog}.}. Furthermore, they do not contribute to the scalar potential that we shall discuss in Section~\ref{sec:dsc_constraints} in the context of the dSC. Notwithstanding the importance of form factors in establishing a non-local behavior of gravity, we would like to understand which values of the three dimensionless combinations
\begin{eqaed}\label{eq:IR_params}
    G\Lambda \, , \quad g_C \, , \quad g_R
\end{eqaed}
are allowed starting from any initial condition, \emph{i.e.}, any perturbation of the asymptotically safe UV fixed point along UV-attractive directions. To this end, we have evaluated numerically these combinations in the IR, implementing the substitution of eq.~\eqref{eq:t_planck_sub}. The following plots highlighting the swampland constraints, the IR limits of asymptotically safe RG trajectories, as well as the final intersection between the allowed regions, will pertain to the $(G\Lambda,g_C,g_R)$ theory space.

To conclude this section, let us collect a few words of caution regarding the one-loop approximation. In general, in the context of gravity, one expects it to be only reliable in the IR, despite the appearance of a UV fixed point outside of the perturbative regime. The methods of the functional RG have been employed, both in earlier~\cite{Benedetti:2009rx,Benedetti:2009gn} and recent~\cite{Knorr:2021slg} efforts, to obtain non-perturbative flow equations in the quadratic truncation, but applying our method to extract the allowed region of parameter space in the IR entails highly involved and unstable numerical analysis. In order to circumvent these obstacles, and address the problem in a more quantitative fashion, a natural first step would entail performing novel FRG computations. The simplest relevant setting would include the most general quadratic truncation coupling the electromagnetic field to gravity, which, while daunting, appears feasible via the methods that have been very recently introduced in~\cite{Knorr:2021slg} to study the purely gravitational sector. In light of these (and other related) issues, in this work we have focused on the one-loop approximation as a proof of principle, with the hope of uncovering some instructive general lessons from the results that we are now about to present.

Finally, let us stress that, although quadratic truncations of the gravitational action are typically associated with a loss of physical unitarity~\cite{Stelle:1977ry}, the Stelle ghost could be a truncation artifact~\cite{Platania:2020knd}. Integrating out quantum fluctuations could lead to well-behaved, unitary scattering amplitudes~\cite{Draper:2020bop,Draper:2020knh}, as explicit computations seem to suggest~\cite{Bonanno:2021squ}. This issue was also discussed within the setting explored in this paper in~\cite{Niedermaier:2009zz, Niedermaier:2010zz}.

\section{Results}\label{sec:results}

Let us now describe in detail our results on the allowed values of physical IR parameters that we have obtained from the calculations outlined in the preceding section, along with the swampland constraints that we have discussed in sect.~\ref{sec:swampland}.

\subsection{Infrared limit of asymptotically safe RG trajectories in one-loop quadratic gravity}\label{sec:IR_space}

\begin{figure}[t!]
\centering\includegraphics[scale=0.5]{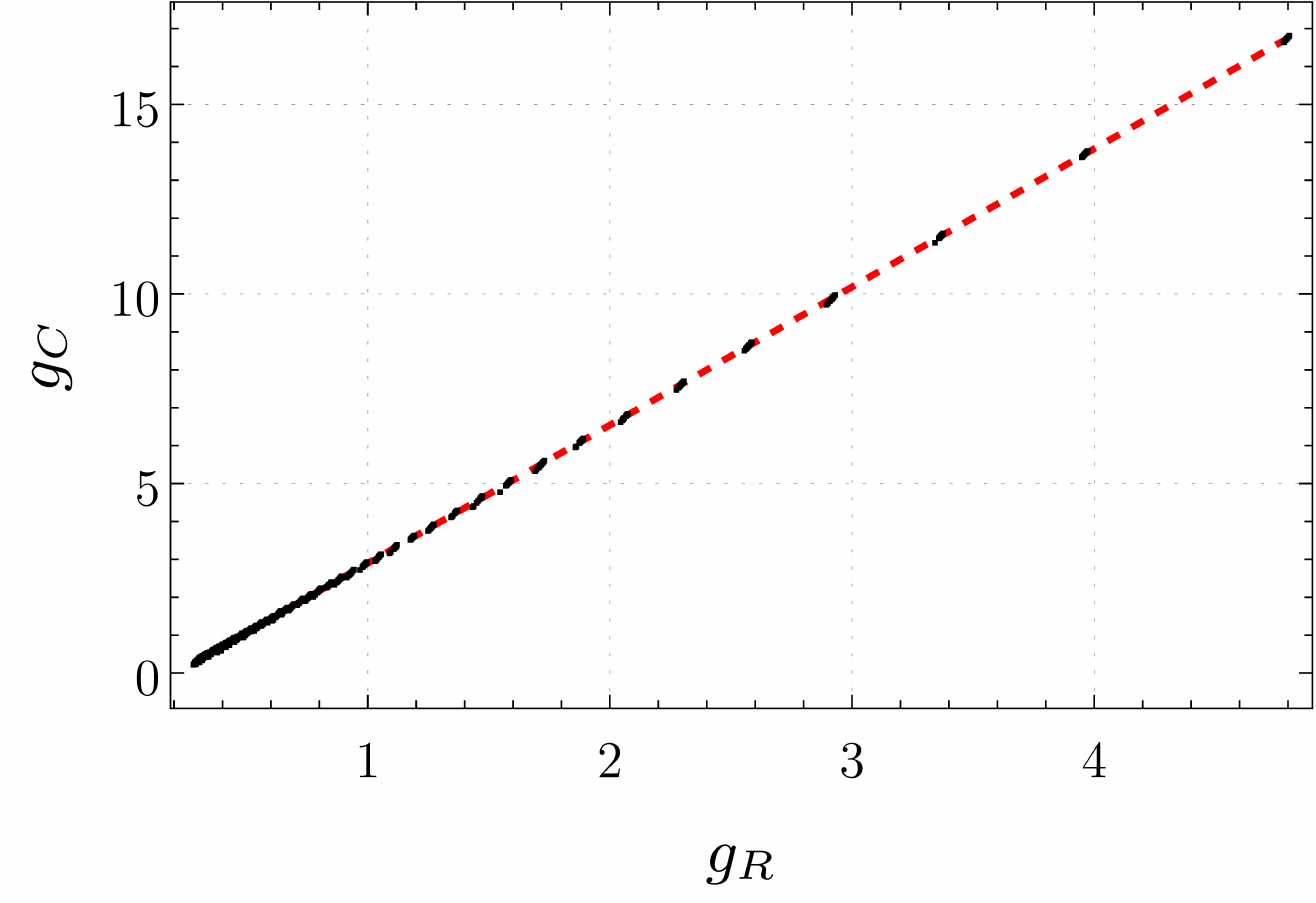}
\caption{The line of equation $g_R = - \, 0.74655 + 3.64447 \, g_C$ fitting the IR values obtained from the flow of eq.~\eqref{eq:betas} sampling UV initial conditions. The covariance matrix evaluates to $\mathcal{O}(10^{-8})$ with~$10^6$ data points.}\label{fig:IR2d-fit}
\end{figure}

\begin{figure}[t!]
$\hspace{-0.2cm}$\includegraphics[scale=0.53]{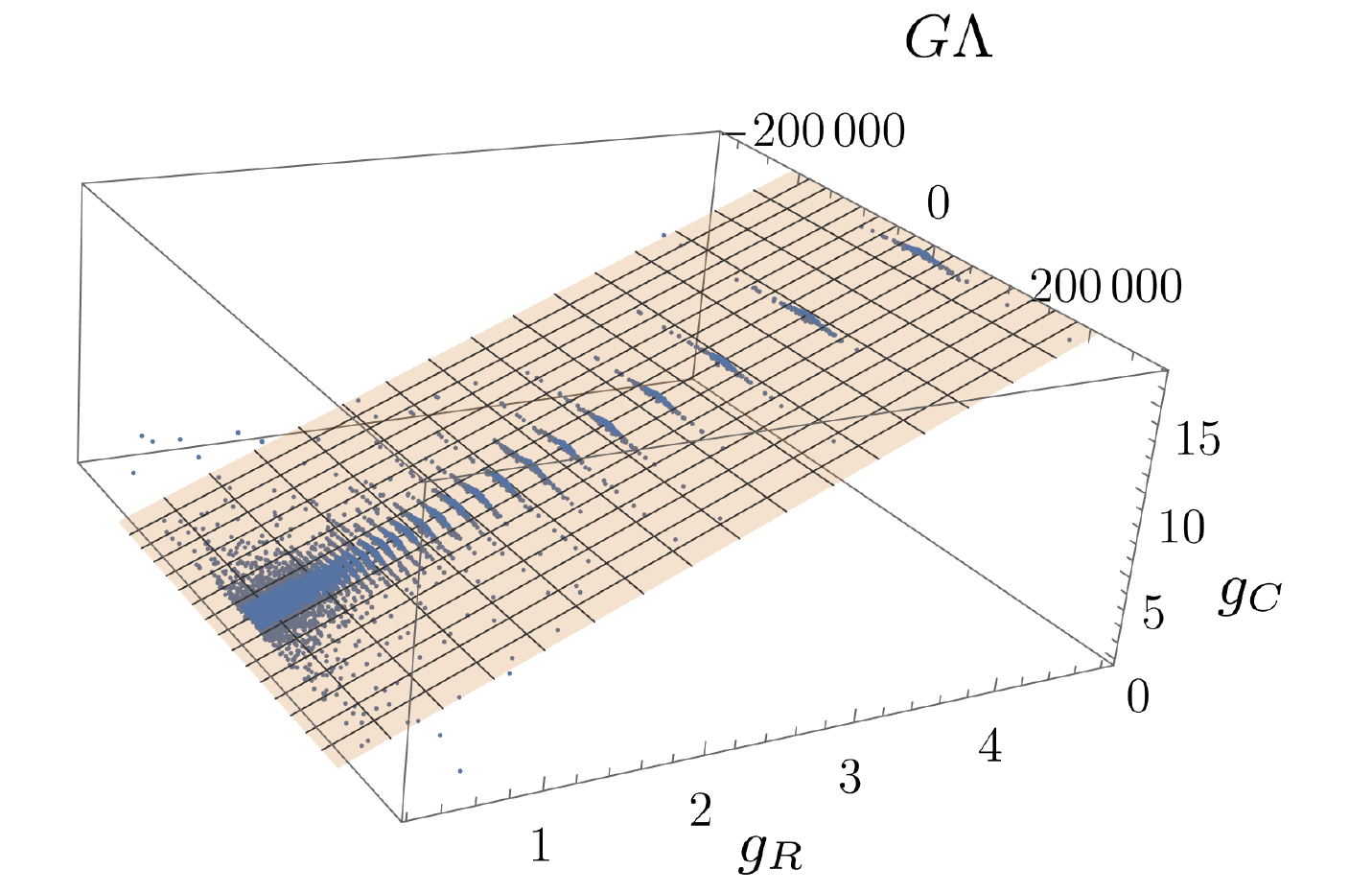}$\hspace{-0.5cm}$\includegraphics[scale=0.53]{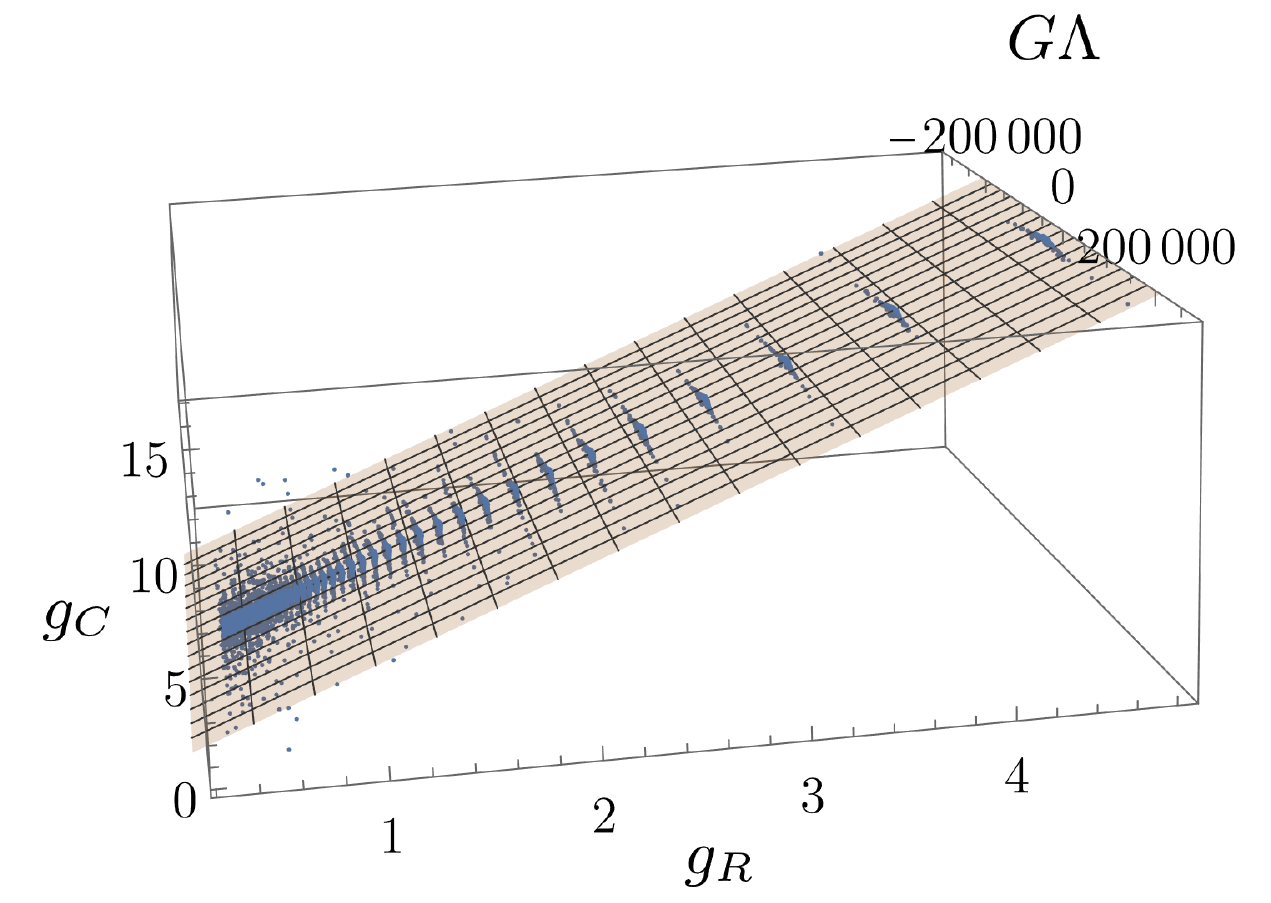}\\
\centering\includegraphics[scale=0.55]{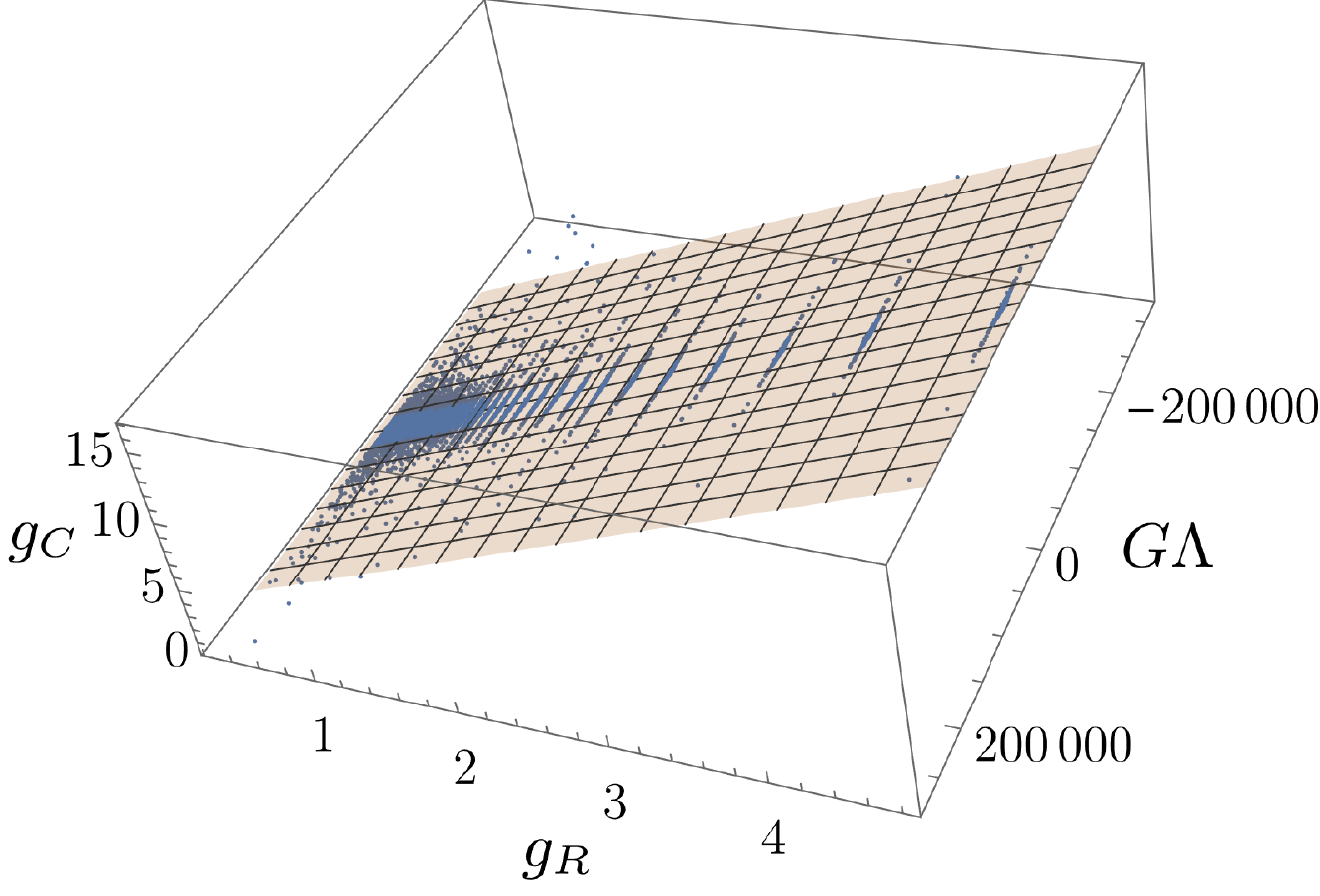}
\caption{Plots depicting the IR endpoints of asymptotically safe RG trajectories. The points lie on the plane of equation $g_R = - \, 0.74655 + 3.64447 \, g_C$. The values of $G\Lambda$ span a vast range, which, together with the closed-form flow of~\cite{Codello:2008vh} depicted in fig.~\ref{fig:flow-reldeformations}, leads us to infer that they are unrestricted.}\label{fig:IR3d}
\end{figure}

In order to uncover the space of physical parameters appearing in the (local sector of the) effective action of eq.~\eqref{eq:form_factor_eft}, we have sampled the space of allowed deformations $(\delta \omega, \delta \lambda, \delta \widetilde{G} , \delta \widetilde{\Lambda})$ from the UV fixed point, and extracted the resulting IR values of the parameters in eq.~\eqref{eq:IR_params} evaluating the flow of $\widetilde{G}$ and $\widetilde{\Lambda}$ for a suitably large RG time $t \approx 30$, exploiting the rapid convergence of the combination in eq.~\eqref{eq:t_planck_sub}. The resulting values for $\widetilde{G}\widetilde{\Lambda} = G\Lambda$, or equivalently $\Lambda/M_\text{Pl}^2$, span a wide range of values, of the order of $10^5$ for the region of initial deformations that we have explored. Moreover, the closed-form flow that one obtains setting the classically marginal couplings to their fixed-point values spans the whole real axis~\cite{Codello:2008vh}. We are thus led to conclude that the allowed (IR) values of $G\Lambda$ are unrestricted. On the other hand, the values of $g_C$ and $g_R$ appear to lie on the line
\begin{eqaed}\label{eq:marginal_line}
    g_R \approx - \, 0.74655 + 3.64447 \, g_C \, ,
\end{eqaed}
as depicted in fig.~\ref{fig:IR2d-fit} and fig.~\ref{fig:IR3d}. This result appears to be very robust upon increase of the sample size, and in particular for $10^6$ points the covariance matrix of the fit is of the order $\mathcal{O}(10^{-8})$. Let us observe that, neglecting the intercept term, eq.~\eqref{eq:marginal_line} follows from eq.~\eqref{eq:log_running} as $t \to t_\text{IR}^-$, whereby $g_R \sim - \, 2\omega_-/3 \, g_C$. Since we instead evaluate the IR couplings at a fixed, albeit sufficiently large, RG time, it is tempting to speculate that the intercept term in eq.~\eqref{eq:marginal_line} is a correction arising from RG trajectories that approach the IR more slowly. Therefore, at least within the scope of our approximations, the presence of a UV fixed point appears to constrain the allowed physical coefficients in eq.~\eqref{eq:IR_params} to a specific plane, and we shall now compare this result to the constraints arising from the swampland conjectures that we have discussed in sect.~\ref{sec:swampland}. 

\subsection{Constraints on quadratic gravity from WGC}\label{sec:wgc_constraints}

As we have discussed in sect.~\ref{sec:wgc_intro}, the WGC entails positivity bounds for the Wilson coefficients of the higher-derivative corrections to Einstein gravity. Since these bounds involve charged particles and black holes, higher-derivative couplings of a $U(1)$ gauge field ought to be included, although the resulting RG flow is extremely involved technically and has not been computed hitherto. On the other hand, the considerations of~\cite{Charles:2017dbr, Charles:2019qqt, Cano:2021tfs}, based on electromagnetic duality, show that one can still make use of our results to constrain higher-derivative corrections in a \emph{duality-invariant} scenario using the WGC. To this end, expressing the higher-curvature in terms of the $c_i$ coefficients of eq.~\eqref{eq:eft_corr}, the (family of) positivity bound(s) of~\cite{Cheung:2018cwt} reads
\begin{eqaed}\label{eq:positivity_bound}
    (1-\xi)^2\,c_0+20\xi c_3-5\xi(1-\xi)(2c_3)>0
\end{eqaed}
where $\xi\equiv\sqrt{1-Q^2/M^2}$ is the extremality parameter of Reissner-Nordstr\"{o}m black holes with mass $M$ and charge $Q$, $0<\xi<1/2$ for black holes with positive specific heat and
\begin{eqaed}\label{eq:c0_coeff}
    c_0 \equiv c_2 + 4 \, c_3 \, .
\end{eqaed}
In terms of the couplings in eq.~\eqref{eq:quadratic_lagrangian}, the bound of eq.~\eqref{eq:positivity_bound} takes the simpler form
\begin{eqaed}\label{eq:simpler_wgc_bound}
    \frac{1}{\lambda}(10\, \theta \, (\xi +1) \xi +6 \xi ^2+3 \xi +1)>0 \, ,
\end{eqaed}
which holds for~$\lambda>0$ provided that~$\xi>0$ (which is always satisfied by the extremality parameter) and that~$\theta>0$. As we have discussed in sect.~\ref{sec:one-loop}, the latter condition is fulfilled  if $\delta \theta > 0$, since $\delta \lambda > 0$ in eq.~\eqref{eq:marginal_flow}. Hence, the (duality-invariant) WGC constrains~$\lambda > 0$, which is included by the analysis of the preceding section and does not entail additional conditions. Let us observe that, although $\theta$ encodes the coupling of the Gauss-Bonnet invariant, it contributes to the entropy of a black hole~\cite{Myers:1988ze,Myers:1998gt,Clunan:2004tb} even in four dimensions, where it is topological. It does not contribute in the limit $\xi \ll 1$, since the resulting bound also describes the positivity of the extremality ratio~\cite{Kats:2006xp,Charles:2019qqt}.

\subsection{Constraints on quadratic gravity from dS and TC conjectures}\label{sec:dsc_constraints}

Let us now discuss the constraints arising from the dSC and the TCC. As we have anticipated in sect.~\ref{sec:dsc_tcc_intro}, we shall focus on the bounds that the dSC and the TCC entail for the scalar potential that arises from the higher-derivative corrections of eq.~\eqref{eq:form_factor_eft}. In order to extract the potential proper, we shall concern ourselves with the local sector of the theory, neglecting the form factors and the Weyl term, which vanishes on cosmological backgrounds.

\begin{figure}[t!]
\centering\includegraphics[scale=0.65]{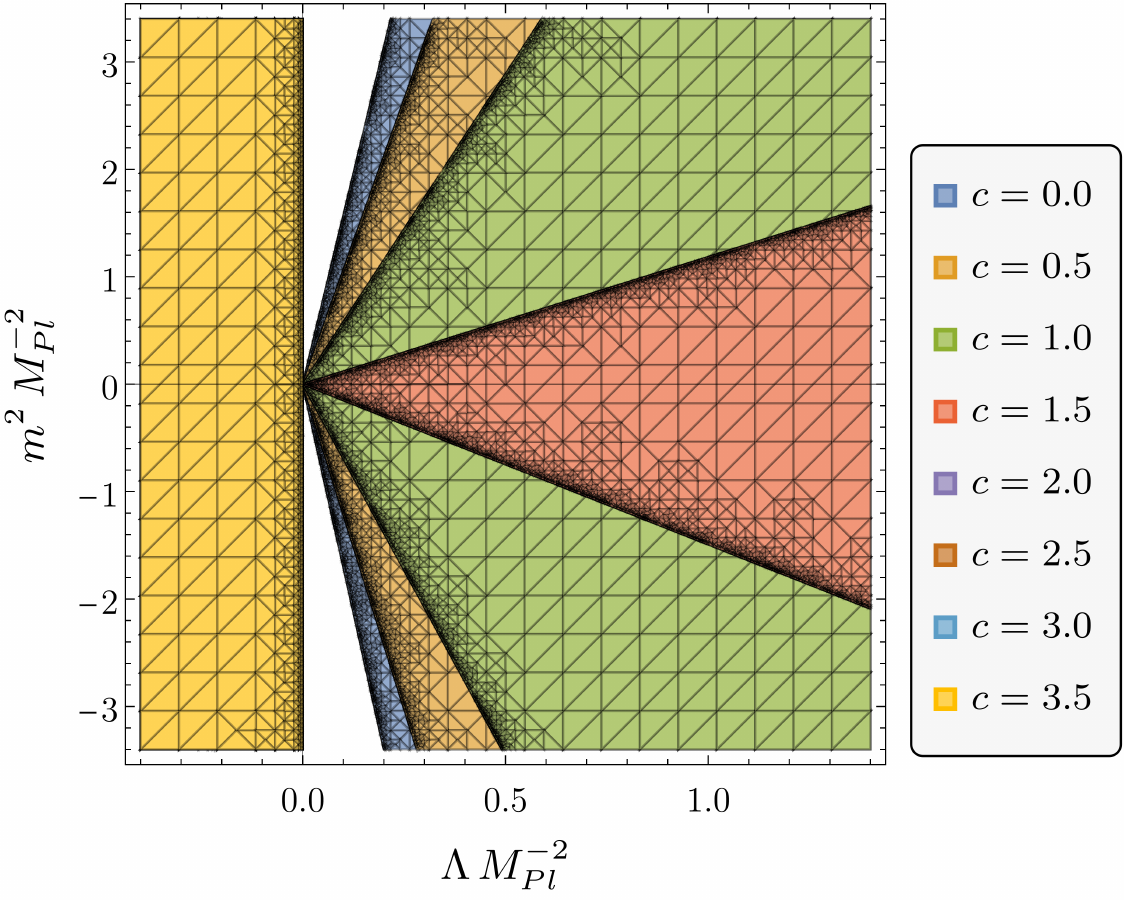}$\,\,$\includegraphics[scale=0.65]{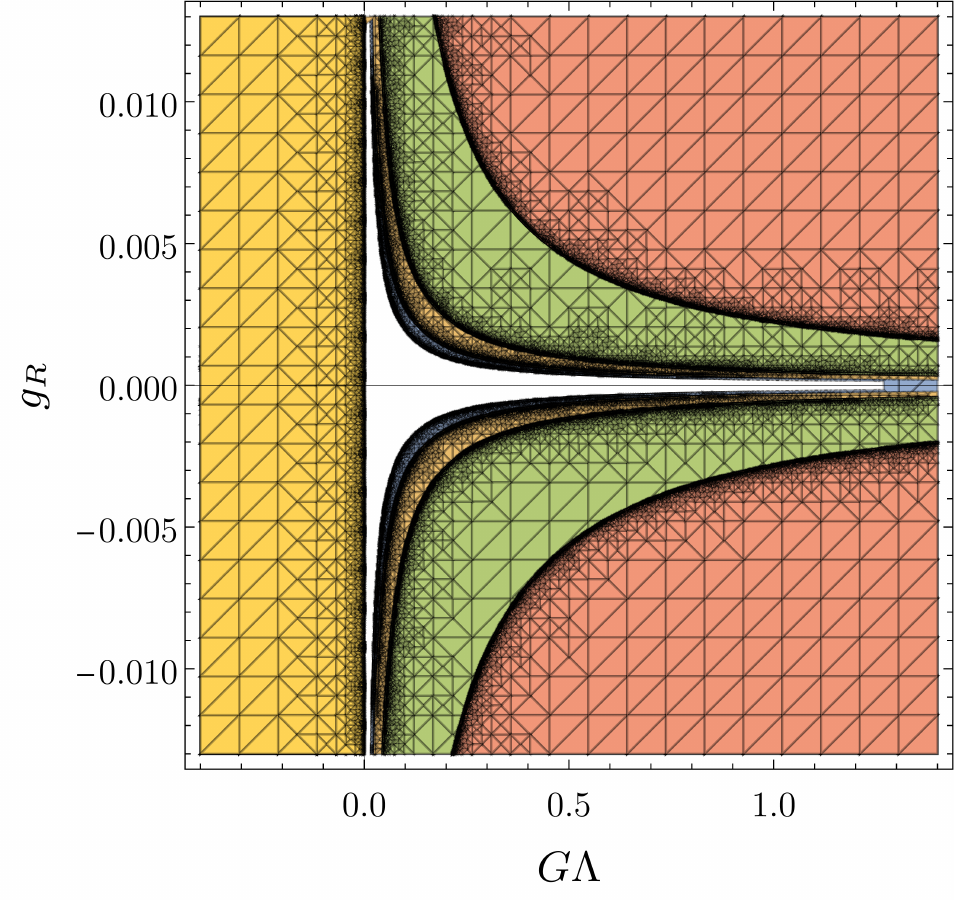}\\
\centering\includegraphics[scale=0.65]{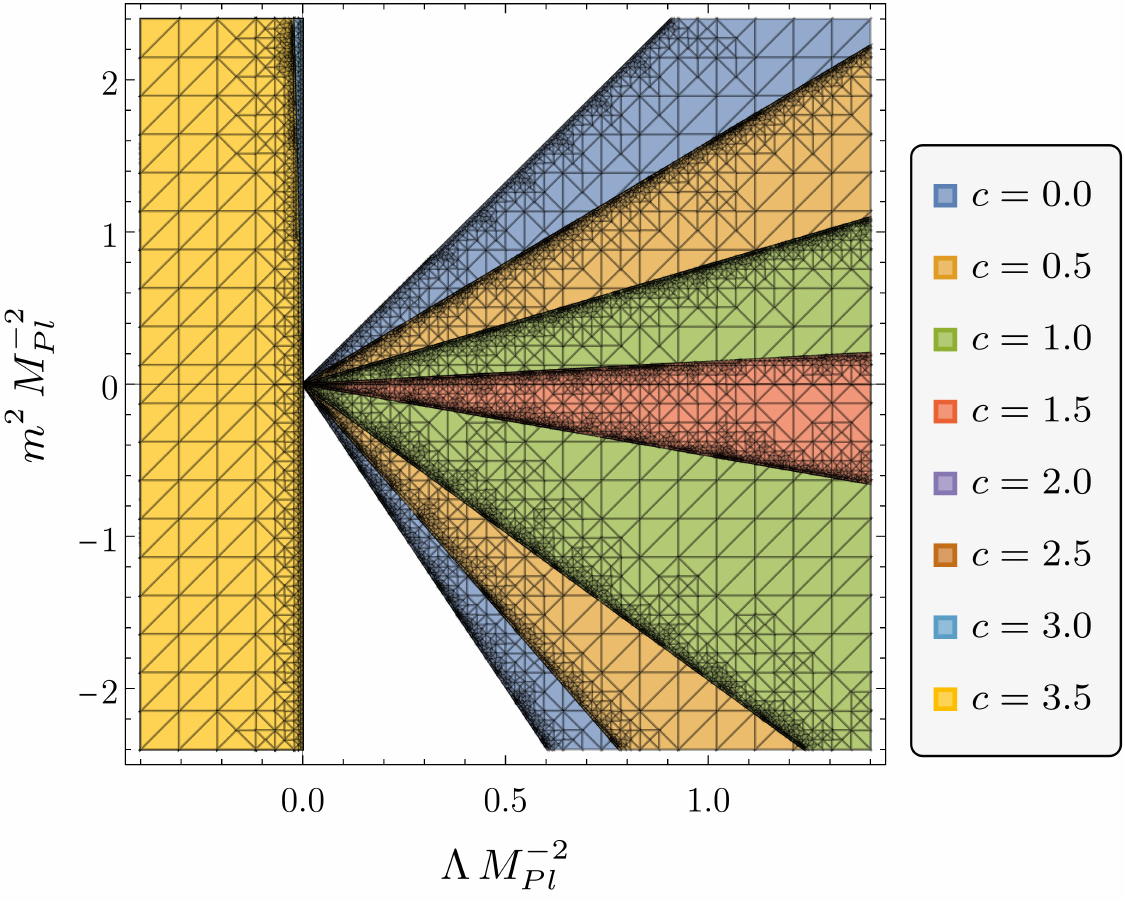}$\,\,$\includegraphics[scale=0.65]{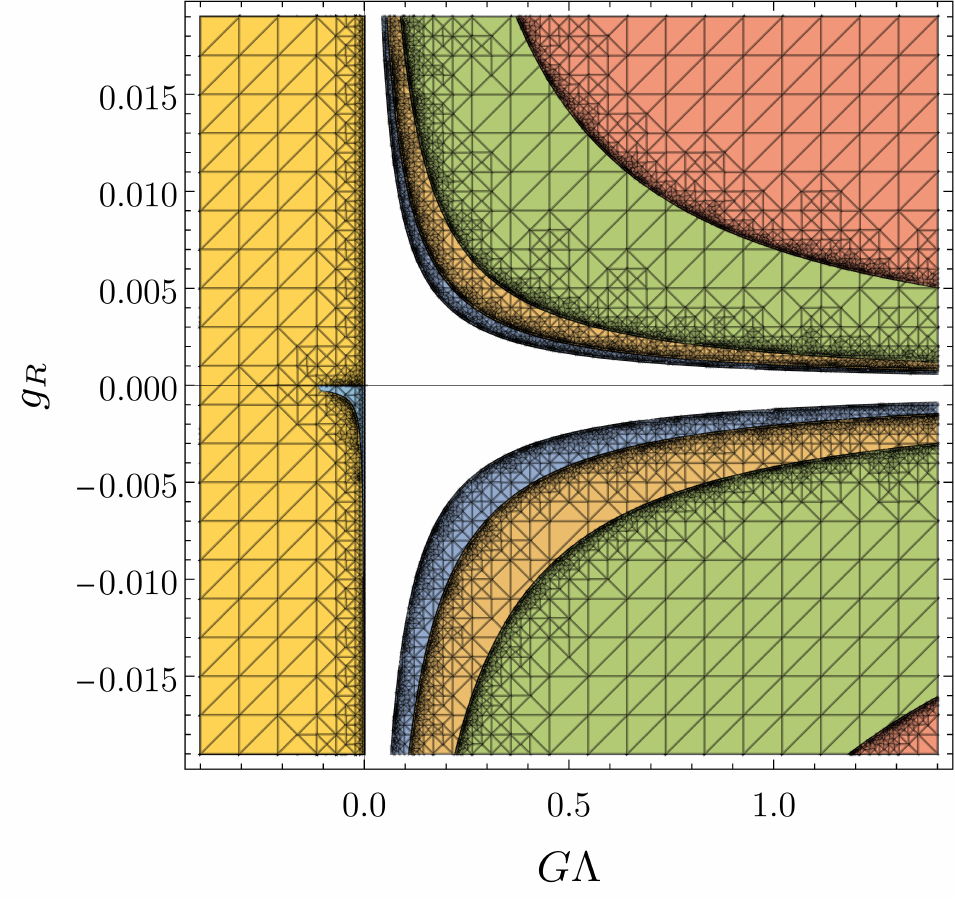}\\
\centering\includegraphics[scale=0.65]{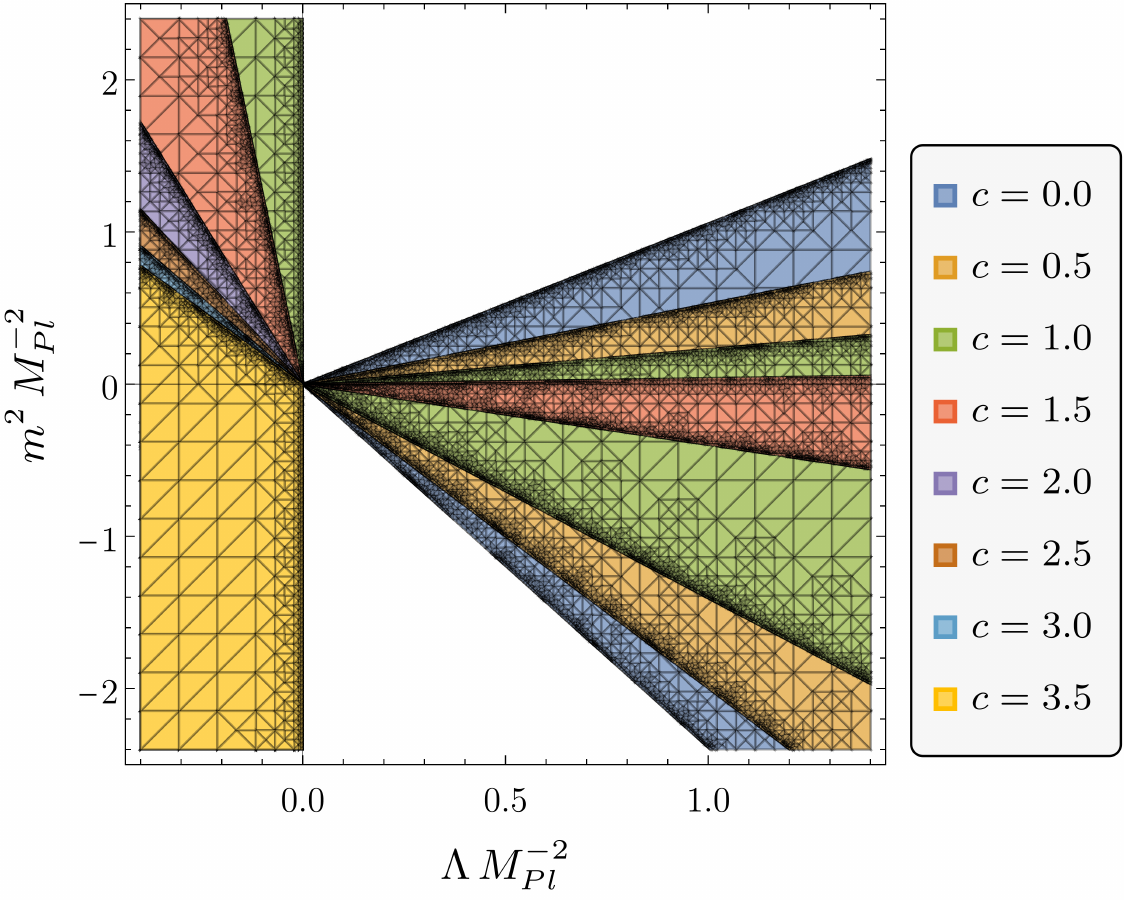}$\,\,$\includegraphics[scale=0.65]{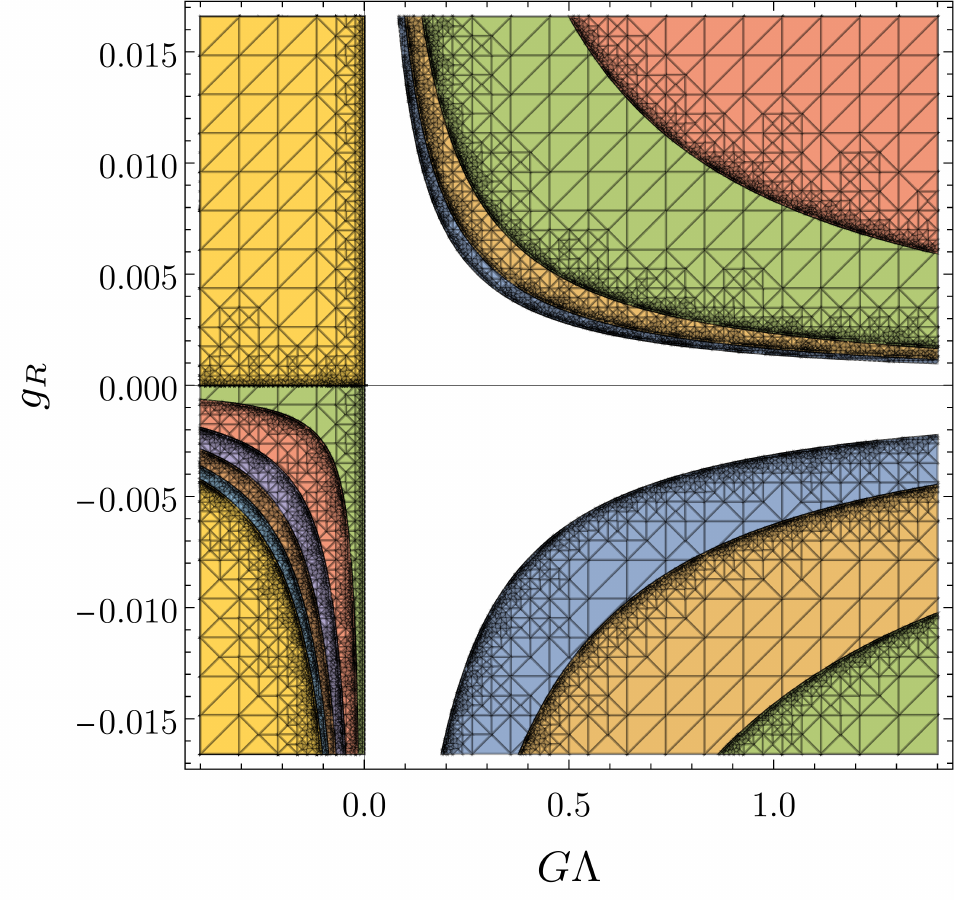}
\caption{dSC/TCC constraints for $f=0.1$ (top panel), $f=0.5$ (central panel) and $f=1$ (bottom panel), and various values of $c$. Due to the inverse relation in eq.~\eqref{eq:grm2} between the inflaton mass in Planck units $m^2/M_{\text{Pl}}^2$ and the coupling $g_R$, the linear bounds in the~$(m^2, \Lambda)$ plane translate into hyperbolas in the~$(g_R,G\Lambda)$ plane. \label{fig:TCC1}}
\end{figure}

\begin{figure}[t!]
\centering\includegraphics[scale=0.65]{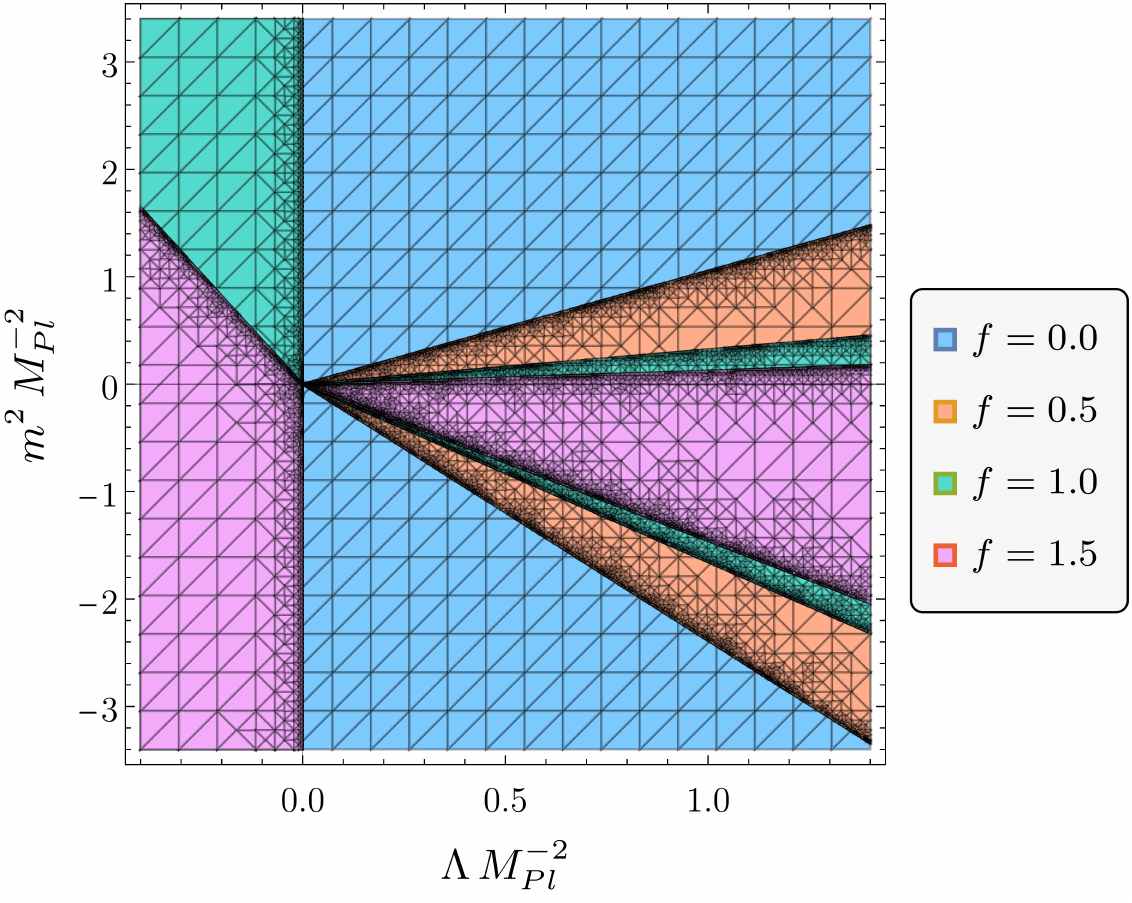}$\,\,$\includegraphics[scale=0.65]{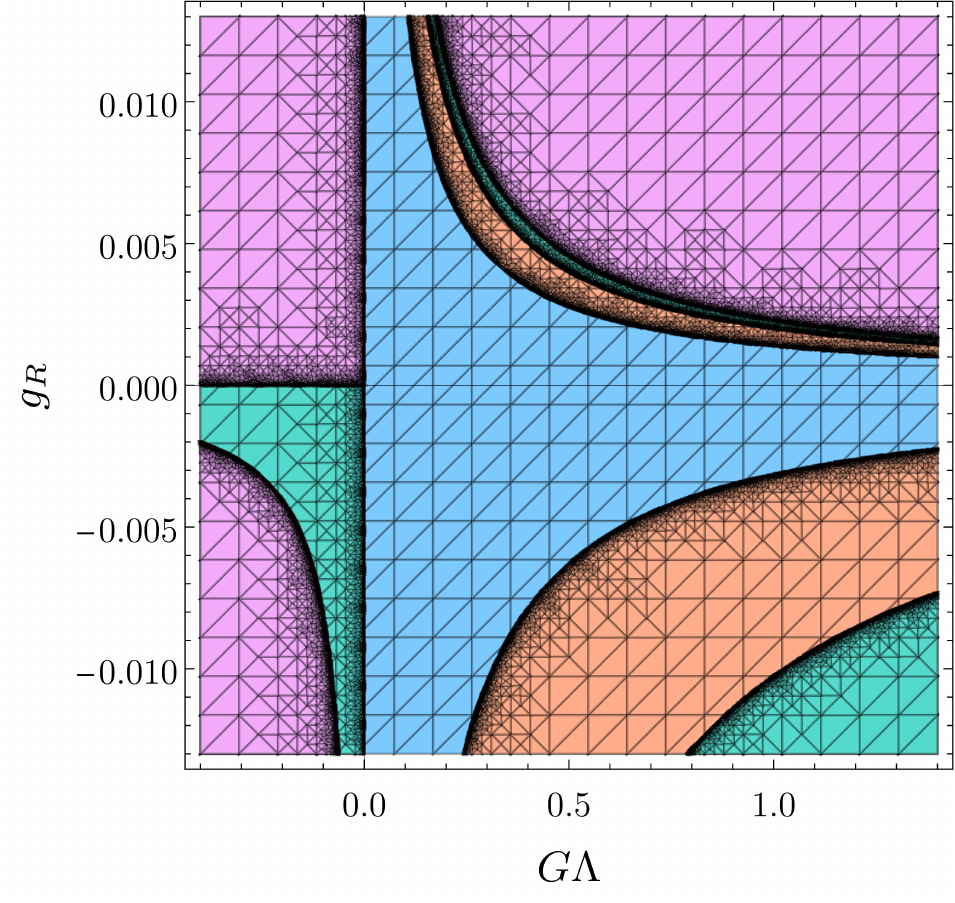}
\caption{TCC constraints, corresponding to $c=\sqrt{2/3}$, for various values of $f$. The bounds are not qualitatively different from to the dSC bounds displayed in fig.~\ref{fig:TCC1}.\label{fig:TCC2}}
\end{figure}

Following the standard procedure to obtain inflaton potentials from $F(R)$ Lagrangians (see, \emph{e.g.},~\cite{inflationaris,Platania:2019qvo}), one begins from $R^2$ gravity with a cosmological constant,
\begin{eqaed}\label{eq:our_f}
    F(R)=\frac{1}{16\pi G}\left(R-2\Lambda + \frac{R^2}{6m^2}\right) \, ,
\end{eqaed}
where the coupling $g_R$ is related to the inflaton mass according to
\begin{eqaed}\label{eq:grm2}
    g_R = - \, \frac{M_\text{Pl}^2}{(8\pi)\cdot 12m^2} \, .
\end{eqaed}
One then arrives at the inflaton potential
\begin{eqaed}\label{eq:scalar_pot}
    V(\phi)=\frac{M_\text{Pl}^2}{8\pi}\,e^{-2\sqrt{\frac{2}{3}}\frac{\phi}{M_\text{Pl}}} \left(\frac{3 m^2}{4}\left(e^{\sqrt{\frac{2}{3}}\frac{\phi}{M_\text{Pl}}}-1\right)^2+\Lambda \right) \, .
\end{eqaed}
In order to retain compatibility with the EFT, we shall consider field values in eq.~\eqref{eq:field_range} around $\phi \ll M_\text{Pl}$, since it corresponds to small curvatures. Indeed, the procedure to obtain inflationary potentials from quadratic gravity yields $\phi = \sqrt{\frac{3}{2}} \, M_\text{Pl} \log\left( 1 + \mathcal{O}(M_\text{Pl}^{-2})\right)$~\cite{inflationaris,Platania:2019qvo}. One can then study dSC and TCC constraints of eqs.~\eqref{eq:dsc_basic},~\eqref{eq:ref_dsc} and~\eqref{eq:tcc_basic} numerically varying the $\mathcal{O}(1)$ constants $c$ and $f$, imposing that the bounds be satisfied for all $\phi$ in the range allowed by eq.~\eqref{eq:field_range}. The resulting regions are highlighted in fig.~\ref{fig:TCC1} and in fig.~\ref{fig:TCC2}, where each panel corresponds to a particular value of $f$ and consists of two plots which display the bounds in the $(m^2,\Lambda)$ (left panels) and $(g_R,G\Lambda)$ (right panels) planes. Due to the inverse relation in eq.~\eqref{eq:grm2} between $m^2/M_{\text{Pl}}^2$ and $g_R$, the linear bounds in the~$(m^2, \Lambda)$ plane translate into hyperbolas in the~$(g_R,G\Lambda)$ plane. Note that whether the dimensionless minimum $\phi_{\mathrm{min}}/M_{\mathrm{Pl}}$, which exists for $\Lambda \,/{m^2} >-3/4$, falls inside the interval $(-f,+f)$ depends on the ratio $\Lambda/m^2$. In particular, the minimum falls outside the interval for $f<|\phi_\mathrm{min}/M_\mathrm{Pl}|$. Consequently, even if $V(\phi_\text{min})$ were to be positive for some $\Lambda$ and $m^2$, this would not necessarily violate the dSC/TCC bounds for fixed values of $f$ and $c$. Consequently, the bounds displayed in fig.~\ref{fig:TCC1} and in fig.~\ref{fig:TCC2} are non-trivially affected by eq.~\eqref{eq:field_range}, and by the specific values of $f$ and $c$. For instance, it is worth noticing that smaller values of $f$ entail smaller field intervals where the dSC/TCC bounds are to be satisfied. Thus, the bounds are less stringent, and the allowed region bigger. Similarly, the bound is more restrictive for higher values of $c$. In particular, fig.~\ref{fig:TCC2} depicts the bounds derived from $c = \sqrt{2/3}$, the value pertaining to the TCC. While our analysis cannot probe the TCC in the large-excursion regime, where it differs substantially from the dSC, the additional considerations of~\cite{Andriot:2020lea, Rudelius:2021oaz} point to a deeper role of this value of $c$ which could manifest itself, in the low-curvature regime at stake, in further investigations of swampland bounds and/or dimensional reduction.


\subsection{Intersections of allowed regions: compatibility of asymptotic safety with dS, TC and WG conjectures}\label{sec:intersections}

\begin{figure}[t!]
\centering\includegraphics[scale=0.5]{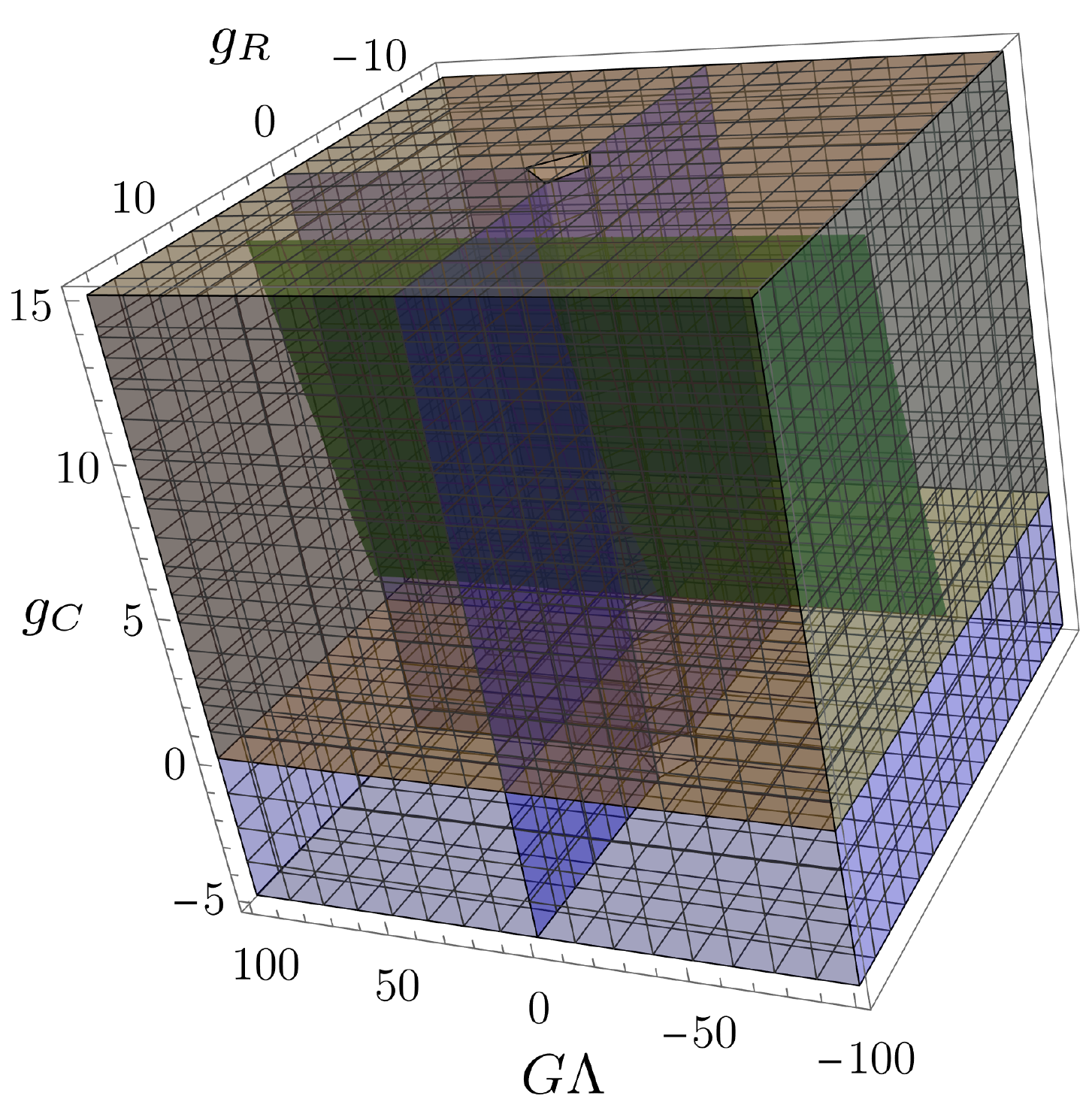}\includegraphics[scale=0.5]{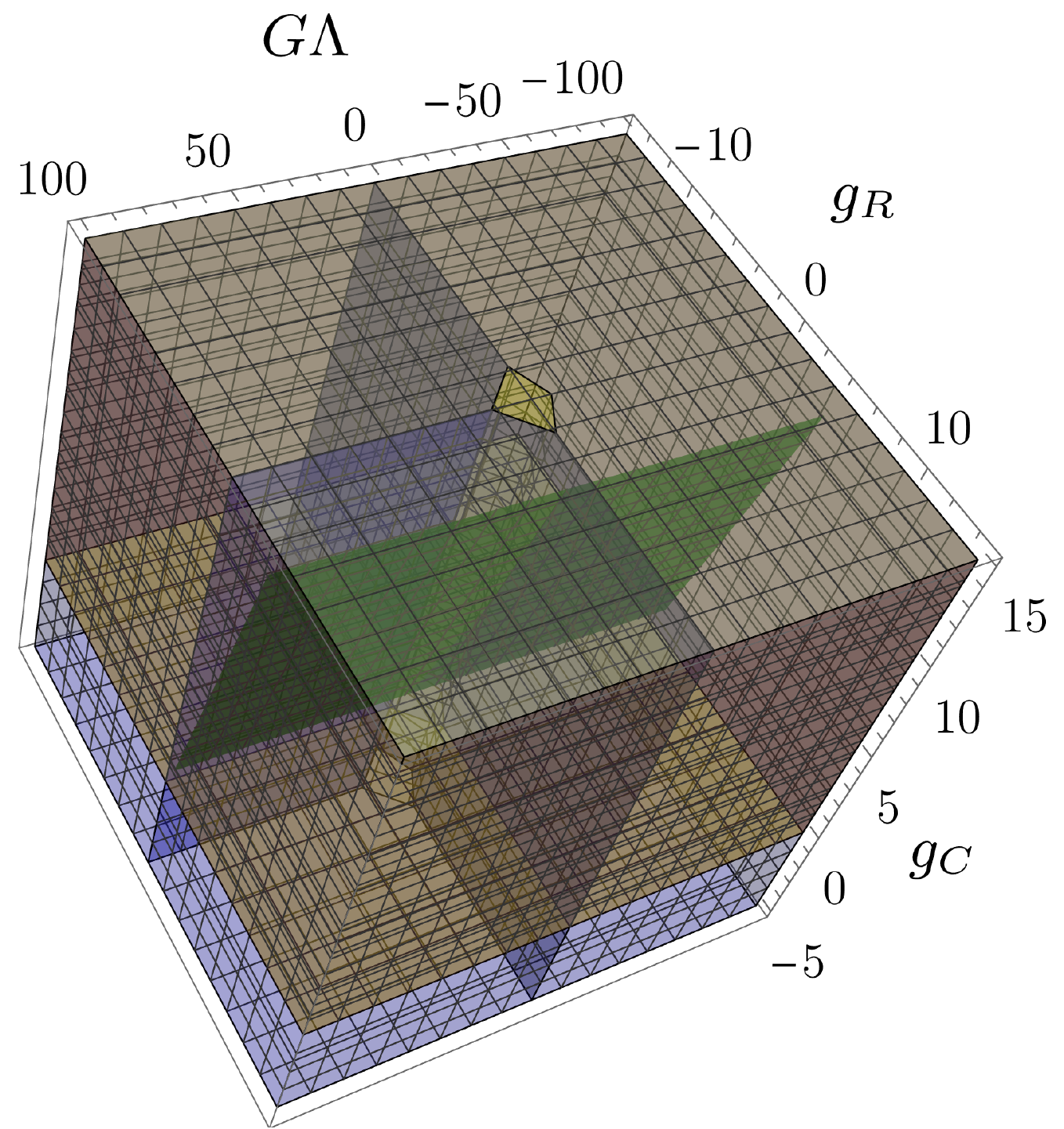}\\\includegraphics[scale=0.5]{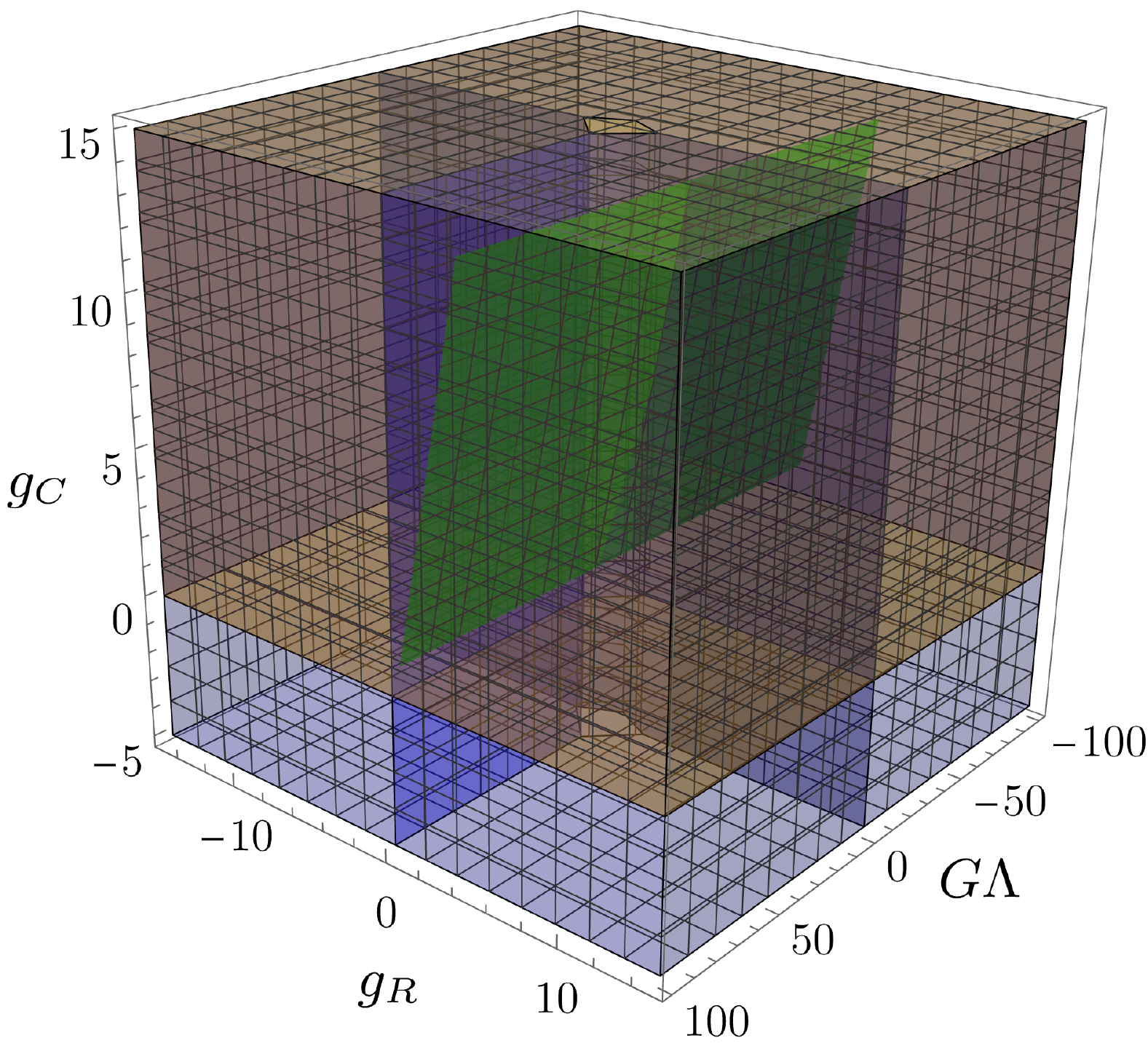}
\caption{Intersections between the regions allowed by asymptotic safety, the WGC and the dSC/TCC for the representative values $c=f=1$. The WGC bound corresponds to the yellow region, while the dSC/TCC bound corresponds to the blue region. The green region depicts the space of IR parameters spanned by asymptotically safe trajectories, which lies within the region allowed by the WGC.}\label{fig:intersections3D}
\end{figure}

We are now ready to collect the results that we have discussed in the preceding sections, and to visualize the intersection between the different allowed regions. Within (an extrapolation of the) one-loop approximation, asymptotic safety of the RG flow constrains the physical IR parameters of eq.~\eqref{eq:IR_params} to lie on the plane of eq.~\eqref{eq:marginal_line}. On the other hand, while the (duality-invariant) WGC does not entail any additional constraint, the dSC/TCC conditions for the inflaton potential place constraints on the cosmological constant and the inflaton mass. One can plot the intersections for any values of the $\mathcal{O}(1)$ constants $c$ and $f$, and fig.~\ref{fig:intersections3D}, the main result of our work, refers to the representative choice $c=f=1$. Albeit difficult to visualize, there is in general a region of the asymptotically safe plane that appears compatible with the swampland constraints that we have investigated. One can straightforwardly verify that the same conclusion is reached for different values of $f$ and $c$. Our findings, based on the quadratic one-loop approximation, thus point at a non-trivial compatibility between the conditions for UV-completeness dictated by asymptotic safety and some of the most relevant swampland conjectures. Consequently, it also points at the possibility, partially supported by~\cite{Basile:2020dzh,Basile:2021krk}, of a connection between the frameworks of asymptotic safety and string theory~\cite{deAlwis:2019aud}. Within this picture, field-theoretical asymptotic safety would serve as a ``pivot'' for the RG flow from string theory to low-energy gravity, in the sense that below a certain scale the flow of string theory toward the IR closely approaches a field-theoretical trajectory controlled by a UV fixed point.

\section{Conclusions}\label{sec:conclusions}

In this paper we have analyzed the intersection of consistency conditions for Wilson coefficients of gravitational EFTs, combining the constraints of asymptotic safety and swampland conjectures. In particular, in sect.~\ref{sec:one-loop} we have employed a systematic method to extract the  hypersurface of allowed IR parameters stemming from UV-complete RG trajectories in gravitational theories by randomly sampling its relevant deformations, and we have applied this technique to the flow equations stemming from one-loop quadratic gravity~\cite{Codello:2008vh}. Despite expecting that this approximation be reliable in the IR, the resulting RG flow exhibits a UV non-Gaussian fixed point, consistently with more refined functional RG computations~\cite{Percacci:2017fkn, Reuter:2019byg, Pawlowski:2020qer}. However, the dimension of its critical surface is larger than what is suggested by the functional RG~\cite{Benedetti:2009rx,Benedetti:2009gn,Falls:2020qhj,Knorr:2021slg}. As we have discussed in detail in sect.~\ref{sec:results}, our findings suggest that the requirement that the RG flow be asymptotically safe constrains the physical parameters to lie on a plane, which we have determined, within our one-loop framework, to a precision of order $\mathcal{O}(10^{-8})$. The values of the cosmological constant in Planck units seems not to be restricted by these considerations, while the classically marginal couplings lie on a line. 

In sect.~\ref{sec:wgc_constraints} and sect.~\ref{sec:dsc_constraints} we have investigated the constraints on the Wilson coefficients arising from the weak gravity conjecture (WGC), the de Sitter conjecture (dSC) and the trans-Planckian censorship conjecture (TCC). In particular, the WGC does not entail additional bounds and is compatible with the UV-complete RG trajectories, while the dSC/TCC bounds are more restrictive. To wit, the Starobinsky-like scalar potential stemming from the (local sector of the) effective action involves the ratio of the cosmological constant to the (squared) inflaton mass, and therefore the corresponding bounds place constraints on the dimensionless ratios $\Lambda/M_\text{Pl}^2$ and $m/M_\text{Pl}$. These bounds depend on some dimensionless $\mathcal{O}(1)$ constants, which we have varied to some extent in our analysis, and generally trace out a region in the plane allowed by asymptotic safety. While we expect that the qualitative results be unaffected by improving the truncation scheme, at least to some extent, it would be interesting to investigate the quantitative deviations in this respect.

While in this work we have focused on the local sector of the effective action, our computation also yields the coefficients of non-local logarithmic form factors, akin to those arising from non-local heat kernel computations~\cite{Barvinsky:1987uw, Barvinsky:1990up, Barvinsky:1990uq, Barvinsky:1993en, Avramidi:1990ap, Codello:2012kq}. It would be interesting to explore their consequences and their role within asymptotically safe gravity~\cite{Knorr:2019atm, Draper:2020bop, Draper:2020knh} and their connection to massive matter fields~\cite{Ohta:2020bsc}.

All in all, our results suggest that swampland constraints can be compatible with restrictions coming from UV completeness of the RG flow, but in a non-trivial fashion: the allowed parameter space is restricted to a non-trivial intersection. In retrospect, one could have expected this result on the grounds that some swampland criteria purport to be necessary conditions for UV completeness that cannot be derived from  purely field-theoretical considerations, and thus they could constrain further the parameter space compatible with asymptotic safety. On the other hand, the one-loop approximation that we have studied already features the appearance of non-local form factors. In general, non-locality at the level of the effective action is a feature of any standard (local) QFT, and thus it is in principle unrelated to possible fundamental non-localities in the bare (fixed-point) action. Precisely how the notion of (non-)locality is realized in quantum gravity is an open and intriguing question, partly related to the problem of observables~\cite{Donnelly:2015hta, Rejzner:2016yuy, Donnelly:2016rvo, Klitgaard:2017ebu, Rudelius:2021azq}. However, a number of semi-classical considerations~\cite{Giddings:1992hh, Susskind:1993if, Almheiri:2012rt, Giddings:2012gc, Dvali:2014ila, Keltner:2015xda, Mann:2015luq} point to the breaking of the familiar concept of locality microscopically. Whether asymptotically safe gravity is realized by a bare action polynomial in derivatives (and thus ``local'' in some sense) is not established yet. Should fundamental non-locality turn out to emerge as a feature of asymptotic safety, this would strengthen its potential connections with the frameworks of non-local gravity~\cite{Modesto:2011kw, Modesto:2017sdr, Buoninfante:2018mre, Buoninfante:2018xiw} and string theory~\cite{Giddings:2006vu}.

Due to the nature of our approximations, this work constitutes only a first step toward determining whether the asymptotic safety scenario is compatible with the peculiar behavior and UV/IR mixing that gravity could exhibit already at the semi-classical level due to black holes (see~\cite{Buoninfante:2021ijy} for a very recent discussion on their validity and limitations), or with general indications from string theory. In particular, a possible connection between asymptotically safe gravity and string theory has been conjectured in~\cite{deAlwis:2019aud}, and it is tempting to speculate that it could explain our findings. Computations combining the functional renormalization group techniques~\cite{Dupuis:2020fhh} with symmetries of string theory~\cite{Veneziano:1991ek, Meissner:1991zj, Meissner:1996sa, Hohm:2015doa, Hohm:2019ccp, Hohm:2019jgu} have provided preliminary evidence in favour of this scenario~\cite{Basile:2020dzh,Basile:2021krk}. This possibility extends to the more general notion of ``effective asymptotic safety''~\cite{Held:2020kze}, and swampland bounds could further constrain which RG flows controlled by the ``effective'' fixed point are closely approached by the RG flow arising from the proper UV completion in the IR.

Most prominently, the absence of continuous global symmetries\footnote{The fate of global discrete symmetries has been investigated in the context of asymptotically safe gravity in~\cite{Eichhorn:2020sbo, Ali:2020znq}.} is supported by a variety of arguments from black-hole physics, string theory and holography~\cite{Misner:1957mt, Banks:1988yz, Kallosh:1995hi, Polchinski:2003bq, Banks:2010zn, Harlow:2018tng, McNamara:2019rup}, and it would be interesting to explore this foundational issue further in the direction that we have outlined in this paper.

\section*{Acknowledgements}

The authors would like to thank F. Saueressig for insighful discussions and B. Knorr for feedback on the manuscript. The authors thank also B. Holdom for spotting a typo.

The work of I.B. is supported by the Fonds de la Recherche Scientifique - FNRS under Grants No. F.4503.20 (“HighSpinSymm”) and T.0022.19 (“Fundamental issues in extended gravitational theories”). A.P. acknowledges support by Perimeter Institute for Theoretical Physics.  Research at Perimeter Institute is supported in part by the Government of Canada through the Department of Innovation, Science and Economic Development and by the Province of Ontario through the Ministry of Colleges and Universities.

\bibliographystyle{JHEP}

\bibliography{AleBib.bib}

\end{document}